\documentclass[11pt]{article}

\usepackage{fullpage}
\usepackage{graphicx}
\usepackage{upref}
\usepackage{enumerate}
\usepackage{latexsym}
\usepackage{pdfpages}
\usepackage[bookmarks]{hyperref}
\usepackage{color,graphics}
\usepackage{comment} 
\usepackage{caption}
\usepackage{subcaption}
\usepackage{wrapfig}
\usepackage{braket}

\usepackage{amssymb}
\usepackage{amsmath}
\usepackage{bbm}
\usepackage{amsthm}
\usepackage{mathrsfs}
\usepackage{mathtools}

\usepackage{epsfig,graphicx}
\usepackage{algorithmic}
\usepackage[ruled,linesnumbered]{algorithm2e}
\usepackage{amsfonts}
\usepackage{tikz} 
\usepackage{varioref}
\usepackage[ansinew]{inputenc}
\usepackage{qcircuit}
\usepackage{authblk}
\usepackage{multirow}
\usepackage{lineno}

\theoremstyle{plain}
\newtheorem{theorem}{Theorem}
\theoremstyle{plain}
\newtheorem{lemma}[theorem]{Lemma}
\theoremstyle{plain}

\theoremstyle{plain}

\theoremstyle{plain}

\theoremstyle{plain}

\theoremstyle{definition}

\newtheorem{definition}[theorem]{Definition}
\theoremstyle{definition}
\newtheorem{fact}[theorem]{Fact}

\theoremstyle{remark}

\theoremstyle{definition}

\AtBeginDocument{}

\DeclareMathOperator{\nat}{\mathbb{N}}

\newcommand{\intg}{\mathbb{Z}}

\newcommand{\tr}{\text{Tr}}

\newcommand{\id}{\mathbb{I}}

\newcommand{\cliff}{\mathcal{C}}
\newcommand{\pauli}{\mathcal{P}}
\newcommand{\clifft}{\mathcal{J}}

\newcommand{\X}{\text{X}}
\newcommand{\Y}{\text{Y}}
\newcommand{\Z}{\text{Z}}
\newcommand{\had}{\text{H}}
\newcommand{\T}{\text{T}}
\newcommand{\CNOT}{\text{CNOT}}
\newcommand{\phase}{\text{S}}

\newcommand{\tof}{\text{TOF}}
\newcommand{\swap}{\text{SWAP}}
\newcommand{\cs}{\text{CS}}

\newcommand{\fred}{\text{FRED}}

\newcommand{\conj}[1]{\overline{#1}}
\newcommand{\vect}[1]{\mathbf{#1}}
\newcommand{\chan}[1]{\widehat{#1}}
\newcommand{\tcount}{\mathcal{T}}
\newcommand{\sde}{\text{sde}}

\newcommand{\gen}{\mathcal{G}}
\newcommand{\cscount}{\mathcal{S}}
\newcommand{\q}[1]{(#1)}

\newcommand{\tofcount}{\mathcal{T}^{of}}

\begin{document}

 \title{Optimizing the non-Clifford-count in unitary synthesis using Reinforcement Learning }

\author[1]{David Kremer \thanks{david.kremer@ibm.com }}
\author[1]{Ali Javadi-Abhari \thanks{ali.javadi@ibm.com }}
 \author[1]{Priyanka Mukhopadhyay \thanks{mukhopadhyay.priyanka@gmail.com, Priyanka.Mukhopadhyay@ibm.com}}
 
 \affil[1]{IBM Quantum, IBM T.J.Watson Research Center, Yorktown Heights, NY 10598}

\date{}
 \maketitle
 
 \begin{abstract}

In this paper we study the potential of using reinforcement learning (RL) in order to synthesize quantum circuits, while optimizing the T-count and CS-count, of unitaries that are exactly implementable by the Clifford+T and Clifford+CS gate sets, respectively. We have designed our RL framework to work with channel representation of unitaries, that enables us to perform matrix operations efficiently, using integers only. We have also incorporated pruning heuristics and a canonicalization of operators, in order to reduce the search complexity. As a result, compared to previous works, we are able to implement significantly larger unitaries, in less time, with much better success rate and improvement factor. Our results for Clifford+T synthesis on two qubit unitaries achieve close-to-optimal decompositions for up to 100 T gates, 5 times more than previous RL algorithms and to the best of our knowledge, the largest instances achieved with any method to date. Our RL algorithm is able to recover previously-known optimal linear complexity algorithm for T-count-optimal decomposition of 1 qubit unitaries. We illustrate significant reduction in the asymptotic T-count estimate of important primitives like controlled cyclic shift (43\%), controlled adder (14.3\%) and multiplier (14\%), without adding any extra ancilla. For 2-qubit Clifford+CS unitaries, our algorithm achieves a linear complexity, something that could only be accomplished by a previous algorithm using $SO(6)$ representation.

 \end{abstract}

 \section{Introduction}
\label{sec:intro}

Quantum computers have shown significant potential to outperform classical computers in certain tasks and over the past few decades a lot of research has been done in order to probe the computational advantage of quantum computers in practically relevant problems like factoring \cite{1994_S}, unstructured database search \cite{1996_G}, etc. Efficient implementations of these quantum algorithms are an absolute necessity in order to realize the claimed theoretical advantages in practice. Analogous to its classical counterpart, quantum algorithms are popularly described and implemented with quantum circuits, which consist of a series of elementary operations or gates belonging to a universal set. Most of the known universal gate sets consist of Clifford group gates and at least one non-Clifford gate. Synthesis and optimization of these circuits are important components of quantum compilation process. For a fault-tolerant implementation it is important to optimize the number of non-Clifford gates like T, CS, Toffoli, etc, since their implementation in most error correction schemes incurs significant overhead in terms of the number of physical qubit requirements, ancilla, measurement operations, etc. Further, these extra components or operations, being themselves error-prone, have the probability of increasing the error rate of the overall encoded circuits. Additionally, the minimum number of non-Clifford gates required to implement certain unitaries is a quantifier of difficulty in many algorithms \cite{2016_BG, 2016_BSS} that try to classically simulate quantum computation. 

An $n$-qubit unitary $W$ can be implemented with a ``discrete finite'' universal gate set (like Clifford+T, V-basis, Clifford+CS, Clifford+Toffoli), such that the unitary $U$ implemented by the circuit is at most a certain distance from $W$ \cite{1997_K, 2006_DN}. A unitary is called exactly implementable by a gate set if there exists a quantum circuit with these gates, that implements it (up to some global phase). Otherwise, it is approximately implementable. \textbf{Unitary synthesis} refers to the broad class of problems that aim to generate a quantum circuit for a given unitary, with additional constraints like optimizing the gate-count, depth, T-count, T-depth, etc. The complexity of any algorithm that synthesizes a quantum circuit for a given $n$-qubit unitary, cannot avoid an exponential dependence on $n$ \cite{2019_AM}. If we impose additional constraints, like synthesizing a circuit with the minimum number of T gates, then the problem becomes harder. Every existing "provable algorithm", for such problems, has an exponential dependence on other factors, like the non-Clifford count i.e. the minimum number of non-Clifford gates required to implement the input unitary. By ``provable algorithms'' we refer to those algorithms that have rigorous proofs about the optimality of their solution and their complexity \cite{2013_AMMR, 2014_GKMR, 2016_RS, 2021_MM, 2022_GMM,  2022_GMM2, 2021_GRT, 2024_Mcs}. Exploiting some special properties of exactly implementable unitaries "heuristic algorithms" have been developed \cite{2021_MM, 2022_GMM, 2024_Mcs} that have a polynomial dependence on the T-count or CS-count. These are algorithms whose claimed optimality of the solutions and complexity depend on some conjectures that are based on certain observations. Nonetheless, with their significantly lower running time and space requirement, it is possible to implement fairly large unitaries on a personal laptop. 

There is an entirely different category of algorithms that also aim to optimize quantum circuits, but a crucial difference is the fact that their input is not a unitary, but an already synthesized circuit of the unitary. We refer to these as re-synthesis algorithms \cite{2014_AMM, 2024_RLBetal}. The quality of their solution depends on the input circuit and no re-synthesis algorithm exists that guarantees an output circuit with the minimum number of T-gates. Their claimed complexity do not account for the cost of synthesizing the input unitary. If we refer to the results of existing papers on re-synthesis algorithms, then we will understand that though these algorithms can be applied to circuits of many qubits, it is practically impossible to synthesize a circuit for an equivalently large unitary by any existing algorithm. Usually, a quantum agorithm or a large unitary is decomposed into smaller unitaries which are synthesized and combined to give the circuit of the larger unitary. If we optimally synthesize these smaller unitaries then we can get better circuits for the larger untiaries. In fact, this can be input to a re-synthesis algorithm and much better overall optimization can be obtained. A small illustration of this fact has been shown in \cite{2021_MM}. 

In summary, non-Clifford-optimal quantum circuit synthesis of unitaries is important and of practical significance, especially from a fault-tolerant perspective. But most existing algorithms are constrained by an impractical running time and space requirement, that scales exponentially with $n$ and the T-count. Hence we probe the usefulness of Artificial Intelligence (AI) in solving this problem. AI tools have been crucial in the advance of many scientific disciplines \cite{2018_DB, 2019_SLGetal, 2021_DSWetal, 2021_JEPetal, 2022_FBHetal, 2023_MBSetal, 2023_WFDetal, 2024_TWLetal}. Recently, techniques of AI have been applied to quantum circuit synthesis and optimization problems \cite{2019_ABIetal, 2019_RO, 2021_AS, 2021_HLZetal, 2021_FNML, 2021_L, 2021_MPRP, 2022_AGOetal, 2022_MZ, 2023_PSFA, 2023_QBW, 2024_FMB, 2024_KVPetal, 2024_RDUetal, 2024_RLBetal, 2025_VGS}. Specifically, in this paper we apply reinforcement learning (RL) in order to synthesize quantum circuits for unitaries that are exactly implementable by the Clifford+T and Clifford+CS gate sets, with an aim to optimize the T-count and CS-count, respectively. Both these gate sets are universal, widely used and well-studied. Additionally, fault-tolerant implementations exist for the non-Clifford T, CS gates \cite{1997_G, 1998_CPMetal, 2009_FSG, 2012_RdCNetal, 2013_PR, 2017_Y, 2018_HH, 2018_HH2, 2020_BCHK}. Most quantum algorithms are implemented with the Clifford+T gate set. Further, due to its natural implementation as an entangling operation in certain superconducting qubit systems whose fidelity is approaching that of single qubit gates \cite{2016_CMBetal, 2016_SMCG, 2020_FNDetal, 2020_GC, 2021_GKLetal, 2021_GRT}, the CS gate has received much attention as a non-Clifford alternative to the T gate.

\subsection{Our results}

We have designed an RL framework that works with the channel represenation of exactly implementable unitaries. Compared to previous papers \cite{2024_FMB, 2024_RDUetal} employing AI tools for unitary synthesis, while optimizing the T-count, we show that we can implement much larger and better optimized (i.e. less non-Clifford count) circuits, and attain a higher success rate within a much shorter time. Further, to the best of our knowledge, no previous work has used AI to optimize the count of a multi-qubit non-Clifford gate. We demonstrate that our algorithm can be used to optimize the CS-count of a unitary. 

There are certain aspects that set our RL algorithm significantly apart from other ML or AI-based unitary synthesis algorithms, for example \cite{2024_FMB, 2024_RDUetal, 2024_KVPetal}. (i) The first major change that we introduce is in the representation of unitaries. Using a transformation, unitaries are represented as array of integers (channel representation). This is in contrast to existing ML or AI algorithms that represent unitaries as array of complex numbers. This simplifies many operations. (ii) Matrix operations like multiplication, inverse, etc can be performed very efficiently using specially designed procedures in this specific representation of unitaries. This helps in significantly reducing the overall time complexity of the RL algorithm. (iii) The next major change is in the re-formulation of the basis gate set. Instead of using Clifford+T or Clifford+CS gate set, we use generating set $\gen_T$ and $\gen_{CS}$, respectively. Any exactly implementable unitary can be written as a product of these generating set unitaries and a Clifford. Further, due to some special properties of these unitaries in the channel representation, it is sufficient for us to work with only the basis set $\gen_T$ and $\gen_{CS}$, having cardinality at most $4^n$ and $O(n^216^{n-2})$, respectively. Previous papers, working with the Clifford+T gate set, work with a basis set of cardinality $O(3^n|\cliff_n|) \in O(3^n 2^{kn^2})$, where $k>2.5$. $\cliff_n$ is the $n$-qubit Clifford group having cardinality $O(2^{kn^2})$ \cite{2014_KS}. This exponential reduction in the basis set size massively reduces the complexity of searching. We are actually able to implement much larger Clifford+T and Clifford+CS circuits. (iv) Another major change that we introduce is in the searching procedure itself. In order to reduce the search space, during both training and testing, we use pruning techniques that depend on some well-defined mathematical properties of the unitaries.  (v) To further reduce the complexity of the searching procedure, we partition unitaries into cosets with a well-defined representative for each coset. It is sufficient to map a given unitary to a coset and then search a circuit for its representative unitary. This canonicalization not only reduces the search time for synthesis (inference) but also allows the model training to reach much higher gate counts with high success rate.

We have compared the performance of our RL algorithm with state-of-the-art T-count-optimal non-ML unitary synthesis algorithm in \cite{2021_MM}. We show that we can implement significantly larger Clifford+T circuits with much less time. We are also able to recover the optimal linear time complexity for the T-count-optimal decomposition of 1-qubit unitaries. We illustrate that we are able to reduce the asymptotic T-count estimate of controlled cyclic shift, controlled adder, integer multiplier circuits by about 43\%, 14.3\% and 14\%, respectively, without introducing extra ancilla. For 2-qubit Clifford+CS unitaries our algorithm achieves a linear time complexity during testing. Such a complexity for this specific case has been accomplished in \cite{2021_GRT}, that works with the $SO(6)$ representation of unitaries.

Here we comment that optimizing the number of generating set unitaries (in $\gen_T$), while synthesizing an exactly implementable unitary, is directly related to optimizing the number of Pauli measurements in Pauli based computation scheme, for example, refer to the recent qLDPC code architecture paper \cite{2025_YSRetal}. This forms another potential application of our algorithms.

\subsection{Relevant works}

 We first review optimal synthesis algorithms that do not use AI tools. Extensive work has been done to synthesize unitaries without optimality constraint \cite{1997_K, 2002_KSVV, 2006_DN, 2011_F, 2020_dBBVA, 2021_MIC, 2002_HRC}. For T-count-optimal synthesis, algorithms have been developed in \cite{2013_KMM, 2014_GKMR, 2016_dMM, 2021_MM} for exactly implementable unitaries, \cite{2013_KMM2, 2015_KMM, 2016_RS} for 1-qubit approximately implementable unitaries, and \cite{2022_GMM2} for arbitrary multi-qubit unitaries. For CS-count-optimal synthesis algorithms have been developed in \cite{2021_GRT} for 2-qubit exactly implementable unitaries and \cite{2024_Mcs} for arbitrary multi-qubit unitaries. Most of the above-mentioned provable algorithms have high complexity and often it is impractical to implement them in practice. So heuristic algorithms have been developed in \cite{2021_MM, 2024_Mcs, 2024_Mtof}. In \cite{2024_PDBV} simulated annealing has been used for T-count-optimal synthesis, while in \cite{2023_LGLS} near-optimal synthesis of 2-qubit unitaries has been done using SO(6) representation.  Work has also been done for T-depth-optimal synthesis \cite{2013_AMMR, 2022_GMM}, V-count-optimal synthesis \cite{2013_BGS, 2015_BBG, 2015_R, 2024_Mv}, Toffoli-count-optimal synthesis \cite{2024_Mtof} and optimization of Clifford gates like CNOT, SWAP \cite{2008_MPH, 2019_LDX, 2020_dBBetal, 2022_BLM, 2022_CSZetal, 2022_GHLetal}. 

Recent years have witnessed a surge in efforts to solve circuit synthesis and optimization problems using techniques from machine learning (ML), ranging from reinforcement learning (RL) to generative models. For example, ML based algorithms for circuit optimization and unitary compilation can be found in \cite{2019_ABIetal, 2019_RO, 2021_FNML, 2021_HLZetal, 2021_L, 2021_MPRP, 2022_AGOetal, 2022_MZ, 2023_QBW, 2024_RLBetal, 2024_FMB, 2024_KVPetal, 2025_DKMetal}. In \cite{2024_RDUetal} RL algorithms have been developed in order to synthesize T-count-optimal circuits for given unitaries.

\subsection{Organization}

Some necessary preliminary definitions and results have been given in Section \ref{sec:prelim}. In Section \ref{sec:results} we have described our implementation results, while in Section \ref{sec:method} we have described our algorithms. Finally we conclude with some discussions in Section \ref{sec:discuss}.
 \section{Preliminaries}
\label{sec:prelim}

In this section we write some necessary definitions and results. The qubits on which a gate acts is mentioned in the subscript with brackets. For example, $X_{\q{q}}$ implies an X gate acting on qubit $q$. $\CNOT_{(i;j)}$ denotes CNOT gate controlled on qubit $i$ and target on qubit $j$. For symmetric multi-qubit gates like CS, where the unitary does not change if we interchange the control and target, we replace the semi-colon with a comma. For convenience, we skip the subscript, when it is clear from the context. More facts about $n$-qubit Cliffords ($\cliff_n$) and Paulis ($\pauli_n$) have been mentioned in Appendix \ref{app:prelim}.

\subsection{Non-Clifford-count of circuits and unitaries}
\label{subsec:countDefn}

\subsubsection*{T-count and CS-count of circuits}

The \emph{T-count and CS-count of a circuit} is, respectively, the number of T-gates and CS-gates in it. 

\subsubsection*{T-count and CS-count of exactly implementable unitaries}

The group generated by the Clifford and T gates corresponds to the set of unitaries exactly implementable by these gates and we denote it by $\clifft_n^T$.  The \emph{T-count} of a unitary $U\in\clifft_n^T$ is the minimum number of T-gates required to implement it (up to a global phase) with a Clifford+T circuit. We denote the T-count of $U$ by $\tcount(U)$. 

The CS-count of unitaries can be defined in an analogous way. The group generated by the Clifford and CS gates is denoted by $\clifft_n^{CS}$.  The \emph{CS-count} of a unitary $U'\in\clifft_n^{CS}$ is denoted by $\cscount(U')$. 

\subsection{Generating set}
\label{subsec:genSet}

Here we define generating sets (modulo Clifford) for unitaries exactly implementable by the Clifford+T and Clifford+CS gate sets. We can express any exactly implementable unitary (up to a global phase) as product of unitaries from these sets and a Clifford. If $P, P_1, P_2\in\pauli_n$, then we define the following unitaries.
\begin{eqnarray}
    R(P) &=&  \frac{1}{2}\left(1+e^{\frac{i\pi}{4}}\right)\id + \frac{1}{2}\left(1-e^{\frac{i\pi}{4}}\right)P   \nonumber\\
    G_{P_1P_2} &=& \left(\frac{3+i}{4}\right)\id+\left(\frac{1-i}{4}\right)(P_1+P_2-P_1P_2) \nonumber
\end{eqnarray}
Now, we define the following set of unitaries.
\begin{eqnarray}
 \gen_T = \left\{R(P) : P\in\pauli_n   \right\}
 \label{eqn:genT}
\end{eqnarray}

 \begin{eqnarray}
    \gen_{CS}=\left\{ G_{P_1P_2} : P_1,P_2\in\pauli_n\setminus\{\id\}; P_1\neq P_2; [P_1,P_2]=0; (P_1,P_2)\equiv (P_2,P_1)\equiv (P_1,\pm P_1P_2)  \right\}
    \label{eqn:genCS}   
\end{eqnarray}
We use $(P_1,P_2)\equiv (P_1',P_2')$ to imply that only one pair is included in the set. We can prove the following about the above-defined sets.
\begin{theorem}
 \begin{enumerate}
  \item \cite{2014_GKMR} Any unitary that is exactly implementable by the Clifford+T gate set can be expressed as,
  \begin{eqnarray}
U=e^{i\phi} \Big(\prod_{j=m}^{1} R(P_j) \Big)C_0
\qquad [R(P_j)\in\gen_T,\quad C_0\in\cliff_n,\quad \phi\in [0,2\pi)].\nonumber
\end{eqnarray}

  \item \cite{2024_Mcs} Any unitary $U'$ that is exactly implementable by the Clifford+CS gate set can be expressed as,
\begin{eqnarray}
    U'=e^{i\phi'} \left( \prod_{j=m'}^1 G_{P_{1j},P_{2j} } \right)C_0' \qquad [G_{P_{1j},P_{2j}} \in\gen_{CS},\quad C_0'\in\cliff_n,\quad \phi'\in [0,2\pi) ]. \nonumber
\end{eqnarray}
 \end{enumerate}
\label{thm:decompose}
\end{theorem}
Hence, we call $\gen_T$ and $\gen_{CS}$ as the \textbf{generating set} (modulo Clifford) for unitaries exactly implementable by the Clifford+T and Clifford+CS gate sets, respectively. We know that $|\gen_T| = 4^n-1$, while  $|\gen_{CS}|\leq \frac{1}{8}(16^n-13^n-4^n+1)+\frac{1}{12}(12^n-2\cdot 6^n) \in O\left(n^216^{n-2}  \right)$ \cite{2024_Mcs}.

\subsection{Channel representation}
\label{subsec:chanRep}

An $n$-qubit unitary $U$ can be completely determined by considering its action on a Pauli $P_s\in\pauli_n : P_s\rightarrow UP_sU^{\dagger}$. The set of all such operators (with $P_s\in\pauli_n$) completely determines $U$ up to a global phase. Since $\pauli_n$ is a basis for the space of all Hermitian $2^n\times 2^n$ matrices we can write
\begin{eqnarray}
    UP_sU^{\dagger}&=&\sum_{P_r\in\pauli_n}\chan{U}_{rs}P_r,\qquad\text{where }\quad \chan{U}_{rs}=\frac{1}{2^n}\tr\left(P_rUP_sU^{\dagger}\right).
    \label{eqn:chanRepDefn}
\end{eqnarray}
This defines a $4^n\times 4^n$ unitary matrix $\chan{U}$ with rows and columns indexed by Paulis $P_r,P_s\in\pauli_n$. We refer to $\chan{U}$ as the \textbf{channel representation} of $U$. If $V = e^{i\phi}U$, for some $\phi\in [0,2\pi)$, then $\chan{V} = \chan{U}$. By Hermitian conjugation each entry of $\chan{U}$ is real. Also, $\chan{UW}=\chan{U}\chan{W}$ and $\left(\chan{U\otimes W}\right)=\chan{U}\otimes\chan{W}$. Since Cliffords map Paulis to Paulis, up to a possible phase factor of -1, so we have the following.
\begin{fact}
    Let $\chan{\cliff_n}=\{\chan{C}:C\in\cliff_n\}$. A unitary $Q\in\chan{\cliff_n}$ if and only if it has one non-zero entry in each row and column, equal to $\pm 1$.
    \label{fact:chanRepCliff}
\end{fact}

\begin{fact}
The channel representation inherits the decomposition from Theorem \ref{thm:decompose} (and in this decomposition there is no global phase factor).
\begin{enumerate}
 \item $
    \chan{U}=\left(\prod_{j=m}^1\chan{R(P_j) }\right)\chan{C_0}\qquad [R(P_j)\in\gen_T,\quad C_0\in\cliff_n]   
$

\item $
    \chan{U'}=\left(\prod_{j=m'}^1\chan{G_{P_{1j},P_{2j}}}\right)\chan{C_0'}\qquad [G_{P_{1j},P_{2j}}\in\gen_{CS},\quad C_0'\in\cliff_n]    
$
\end{enumerate}
    \label{fact:decompose}
\end{fact}
In point (1) of the above fact, if $m=\tcount(U)$ then we have a \textbf{T-count-optimal decomposition} of $U$. Similarly if $m'=\cscount(U')$  in point (2) of the above fact, then we have a \textbf{CS-count-optimal decomposition} of $U'$.

The channel representation of the unitaries in the generating sets have some nice features that facilitate more efficient computation. We briefly describe some of them in the following points. More explicit descriptions can be found in \cite{2021_MM} (Clifford+T) and \cite{2024_Mcs} (Clifford+CS).
\begin{enumerate}
\item The unitaries $\chan{R(P)}$ and $\chan{G_{P_1,P_2}}$, where $R(P)\in\gen_T$ and $G_{P_1,P_2}\in\gen_{CS}$, have entries in $\left\{0,1,\pm\frac{1}{\sqrt{2}}\right\}$ and $\left\{0,1,\pm\frac{1}{2}\right\}$, respectively. This implies that if $U\in\clifft_n^T$, then entries of $\chan{U}$ belong to the ring $\intg\left[\frac{1}{\sqrt{2}}\right]$, while if $U'\in\clifft_n^{CS}$, then entries of $\chan{U'}$ belong to the ring $\intg\left[\frac{1}{2}\right]$.

\item Special data structures have been designed that enable the storage of these unitaries in a much more compact manner, using only integers. For example, suppose $U \in\clifft_n^T$. Then from the previous point, entries of $\chan{U}$ are of the form $\frac{a+b\sqrt{2}}{\sqrt{2}^k}$, where $a,b\in\intg$ and $k\in\nat$. Then we can store $\chan{U}[r,s]$ as a tuple $(a,b,k)$ of integers.

Similarly, let $U'\in\clifft_n^{CS}$. Then entries of $\chan{U'}$ are of the form $\frac{a}{2^k}$, where $a\in\intg$ and $k\in\nat$. So, we can store $\chan{U}[r,s]$ as a tuple $(a,k)$ of integers.

 \item It follows that matrix operations like addition, multiplication, inverse can be done with integer arithmetic.
 
 \item Algorithms have been developed that perform these matrix operations much more efficiently. Specifically, suppose $U_p = \chan{R(P)}U$ and $U_p' = \chan{G_{P_1,P_2}}U'$. Then $U_p$ and $U_p'$ can be computed in time $O\left(2^{4n-1}\right)$ and $O\left(3\cdot 2^{4n-2}\right)$, respectively. The fastest algorithm to multiply two $2^{2n}\times 2^{2n}$ matrices has a time complexity of $2^{4.745278n}$ \cite{2014_lG}. This modest asymptotic improvement in the complexity has a significant impact on the actual running time, especially when we need to perform a lot of such matrix multiplications. For completion, we have described these procedures briefly in Appendix \ref{app:mult}.
\end{enumerate}

From the fact in point (1) we define the following.
\begin{definition}
\begin{enumerate}
 \item For any non-zero $v\in\intg\left[\frac{1}{2}\right]$ the \textbf{smallest 2-denominator exponent}, denoted by $\sde_2$, is the smallest $k\in\nat$ for which $v=\frac{a}{2^k}$, where $a\in 2\intg+1$. We define $\sde_2(0)=0$. For a $d_1\times d_2$ matrix $M$ with entries over this ring we define
\begin{eqnarray}
    \sde_2(M)&=&\max_{a\in [d_1],b\in [d_2]}\sde_2(M_{ab}).  \label{eqn:sde2M}
\end{eqnarray}
 
 \item For any non-zero $v\in\intg\left[\frac{1}{\sqrt{2}}\right]$ the \textbf{smallest $\sqrt{2}$-denominator exponent}, denoted by $\sde_{\sqrt{2}}$, is the smallest $k\in\nat$ for which $v=\frac{a+b\sqrt{2}}{\sqrt{2}^k}$, where $a\in 2\intg+1$. We define $\sde_{\sqrt{2}}(0)=0$. For a $d_1\times d_2$ matrix $M$ with entries over this ring we define
\begin{eqnarray}
    \sde_{\sqrt{2}}(M)&=&\max_{a\in [d_1],b\in [d_2]}\sde_{\sqrt{2}}(M_{ab}).  \label{eqn:sdeSqrt2M}
\end{eqnarray}
 \end{enumerate}
    \label{defn:sde}
\end{definition}
\footnote{In previous works like \cite{2014_GKMR, 2021_MM}, $\sde_{\sqrt{2}}$ has been referred to as simply $\sde$. We have added the subscript in order to differentiate it from $\sde_2$.} The following observations about the change in sde, have been made in previous works.
\begin{lemma}
\begin{enumerate}
 \item Let $U_p=\chan{R(P)}\chan{U}$, where $\chan{U}=\prod_j\chan{R(P_j) }\chan{C}$  and $P, P_j\in\pauli_n$, $C\in\cliff_n$. Then $\sde_{\sqrt{2}}(U_p)-\sde_{\sqrt{2}}(\chan{U})\in\{\pm 1,0\}$ \cite{2021_MM}.
 
 \item Let $U_p'=\chan{G_{P_1,P_2}}\chan{U'}$, where $\chan{U'}=\prod_j\chan{G_{P_{1j},P_{2j} } }\chan{C'}$  and $P_1,P_2,P_{1j},P_{2j}\in\pauli_n$, $C'\in\cliff_n$. Then $\sde_2(U_p')-\sde_2(\chan{U'})\in\{\pm 1,0\}$ \cite{2024_Mcs}.
 \end{enumerate}
    \label{lem:sdeChangeMat}
\end{lemma}

\subsection{Reinforcement Learning (RL)}
\label{subsec:RL}

RL addresses the question of how an autonomous \emph{agent} that senses and acts in its \emph{environment} can learn to choose optimal actions to achieve its goals. Each time the agent performs an action in its environment, a trainer may provide a reward or penalty to indicate the desirability of the resulting state. Formally, RL can be described as a Markov decision process (MDP), which consists of the following.
\begin{itemize}
 \item A set of \emph{states} $\mathcal{S}$, plus a distribution of starting states $p(s_0)$.
 
 \item A set of \emph{actions} $\mathcal{A}$.
 
 \item \emph{Transition dynamics} $p\left(\vect{s}_{t+1} | \vect{s}_t,\vect{a}_t\right)$ that map a state-action pair at time $t$ onto a distribution of states at time $t+1$.
 
 \item An  immediate \emph{reward function} $r\left(\vect{s}_t, \vect{a}_t, \vect{s}_{t+1}\right)$.
 
 \item A \emph{discount factor} $\gamma\in [0,1]$, where lower values place more emphasis on immediate rewards.
\end{itemize}
At each discrete time step $t$, the agent senses the current state $\vect{s}_t$ and interacts with the environment by performing a current action $\vect{a}_t$. Both the agent and the environment then transitions to a new state $\vect{s}_{t+1}$ determined by the transition probability $p\left(\vect{s}_{t+1} | \vect{s}_t,\vect{a}_t\right)$. The environment also responds by giving the agent a reward $\vect{r}_{t+1}=r\left(\vect{s}_t, \vect{a}_t, \vect{s}_{t+1}\right)$. The goal of the agent is to learn a \emph{policy} that maximises the expected \emph{return}.

In general, the policy $\pi$ is a mapping from states to a probability distribution over actions, i.e. $\pi :\mathcal{S}\rightarrow p(\mathcal{A}=\vect{a}|\mathcal{S})$. If the MDP is \emph{episodic}, i.e. the state is reset after each episode of length $T$, then the sequence of states, actions and rewards in an episode constitute a \emph{trajectory} or \emph{rollout} of the policy. Every rollout of a policy accumulates rewards from the environment, resulting in the return $R = \sum_{t=0}^{T-1}\gamma^t\vect{r}_{t+1}$. The goal of RL is to find an optimal policy, $\pi^*$, which achieves the maximum expected return from all states.
\begin{eqnarray}
 \pi^* = \arg\max_{\pi} \mathbb{E}[R|\pi]   \nonumber
\end{eqnarray}
It is also possible to consider non-episodic MDPs, where $T=\infty$. In this situation $\gamma<1$ prevents an infinite sum of rewards from being accumulated. Given a policy $\pi$ and a general state $\vect{s}$, we denote the expected return of $\vect{s}$ as $v_{\pi}(\vect{s})$, also known as the \emph{value} function. The expected return of taking an action $\vect{a}$ in a state $\vect{s}$ under $\pi$, denoted $q_{\pi}(\vect{s}, \vect{a})$, is referred to as the \emph{state-action} or \emph{Q-value} function.
 \section{Results}
\label{sec:results}

In this section we describe our implementation results. We have implemented our RL algorithm in order to synthesize 1 - 4 qubit unitaries exactly implementable by the Clifford+T and Clifford+CS gate sets, while optimizing the T-count and CS-count, respectively. The algorithms have been implemented on an Intel(R) Xeon(R) Gold 6130 CPU at 2.10 GHz, with 56 cores, 256 GB RAM and Nvidia A100 80 GB GPU. For the training, the GPU and 32 cores of CPU have been used. For the benchmarking, only a single core of CPU (no GPU) has been used. 

For each universal gate set, we have generated some random unitaries in the following way. We randomly sample some generating set unitaries of that particular gate set. We multiply their channel representations using the algorithms described in Appendix \ref{app:mult}, that work with integer arithmetic and are very efficient. Multiplication by a trailing Clifford operator can be realized by randomly permuting the columns and multiplying each  column (i.e. all its entries) by $-1$ with probability $\frac{1}{2}$ (for example refer to Algorithm 12 in Supplementary Information of \cite{2024_Mcs}). These randomly generated unitaries are the input to our RL algorithms, with which we synthesize a circuit for these unitaries. We emphasize that the output of our algorithms are a sequence of generating set unitaries, each of which has non-Clifford count 1. Each of these unitary can be efficiently implemented with the corresponding universal gate sets, as discussed in Appendix \ref{app:cktGen}. The complexity of synthesizing each generating set unitary is $O(n^2)$, where $n$ is the number of qubits. The trailing Clifford can also be efficiently generated with complexity polynomial in $n$, using for example, the algorithms in \cite{2004_AG, 2022_BLM, 2024_KVPetal}. We do not synthesize each of these generating set unitaries and the trailing Clifford explicitly, since we focus solely on getting the non-Clifford count. Here we also remark that the output non-Clifford count or the number of generating set unitaries output by our algorithms does not depend on multiplication by a Clifford (as should be the case). This is because all matrix operations involve row additions or subtractions and thus permuting the columns do not change the sde values, which play the determining factor in our algorithms. It merely changes the signs and hence are not important for our purpose.

As mentioned earlier, the problems of T-count-optimal and CS-count-optimal synthesis of unitaries are hard and the complexity of the provably optimal synthesis algorithms, in general, depend exponentially on the number of qubits, T-count \cite{2014_GKMR, 2021_MM} and CS-count \cite{2024_Mcs}. It is not practical to implement reasonably large unitaries with these algorithms and hence we cannot test if the output circuit generated by our algorithms are optimal. Instead, we do the following. For each input unitary we know at least one circuit and its non-Clifford count. More specifically, we know that each generating set unitary has 1 non-Clifford gate and so the number of generating set unitaries sampled while randomly synthesizing the input unitary gives the number of non-Clifford gates in one circuit of the corresponding unitary. This we refer to as the "input non-Clifford count". The number of generating set unitaries output by our RL algorithms gives the "output non-Clifford count", that is the number of non-Clifford gates in the output circuit. 

\begin{figure}[htb]
    \begin{subfigure}[b]{0.3\textwidth}
    \centering
    \includegraphics[width=\textwidth]{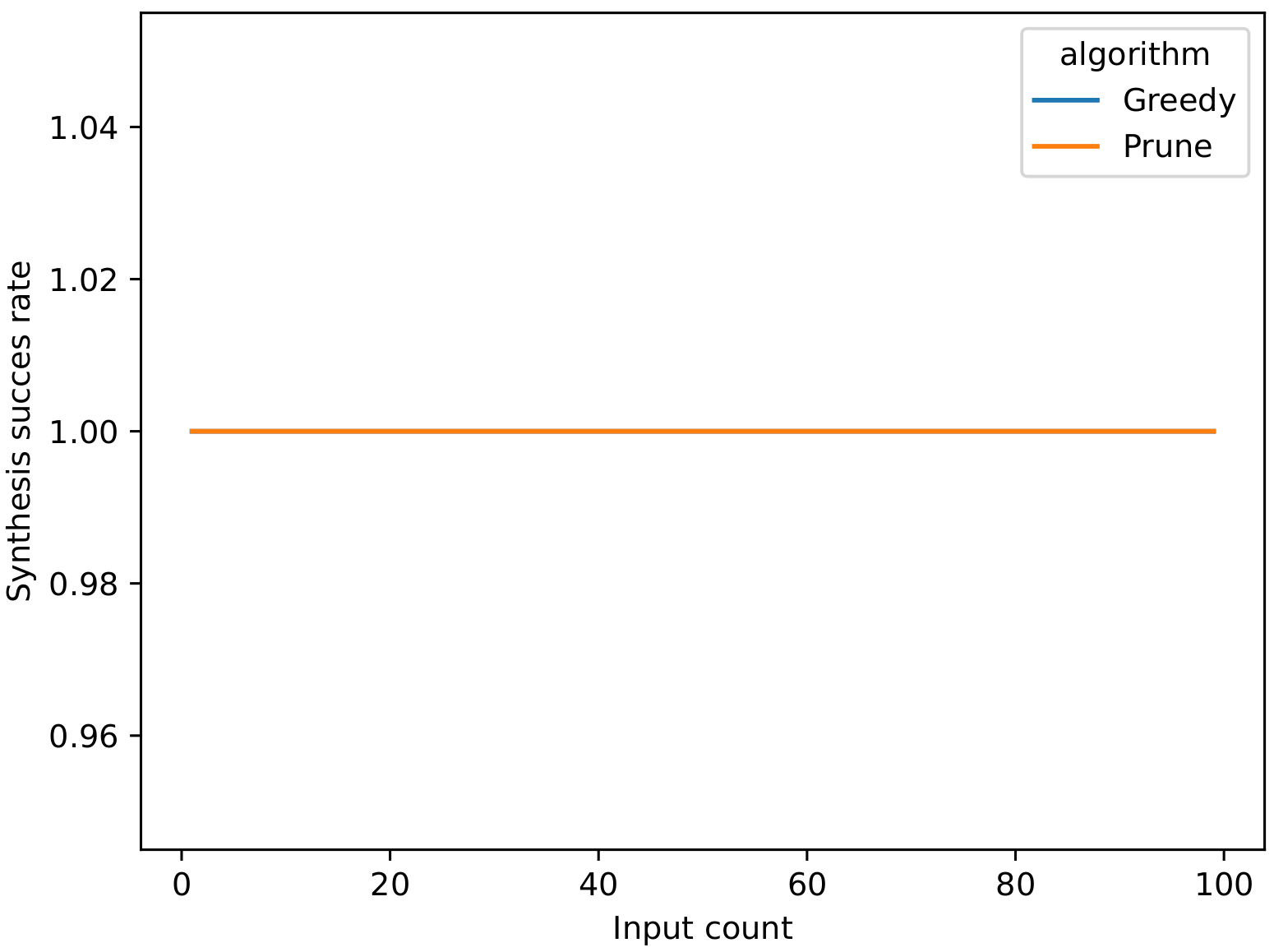}    
    \caption{}
    \end{subfigure}
\hfill
    \begin{subfigure}[b]{0.33\textwidth}
    \centering
    \includegraphics[width=\textwidth]{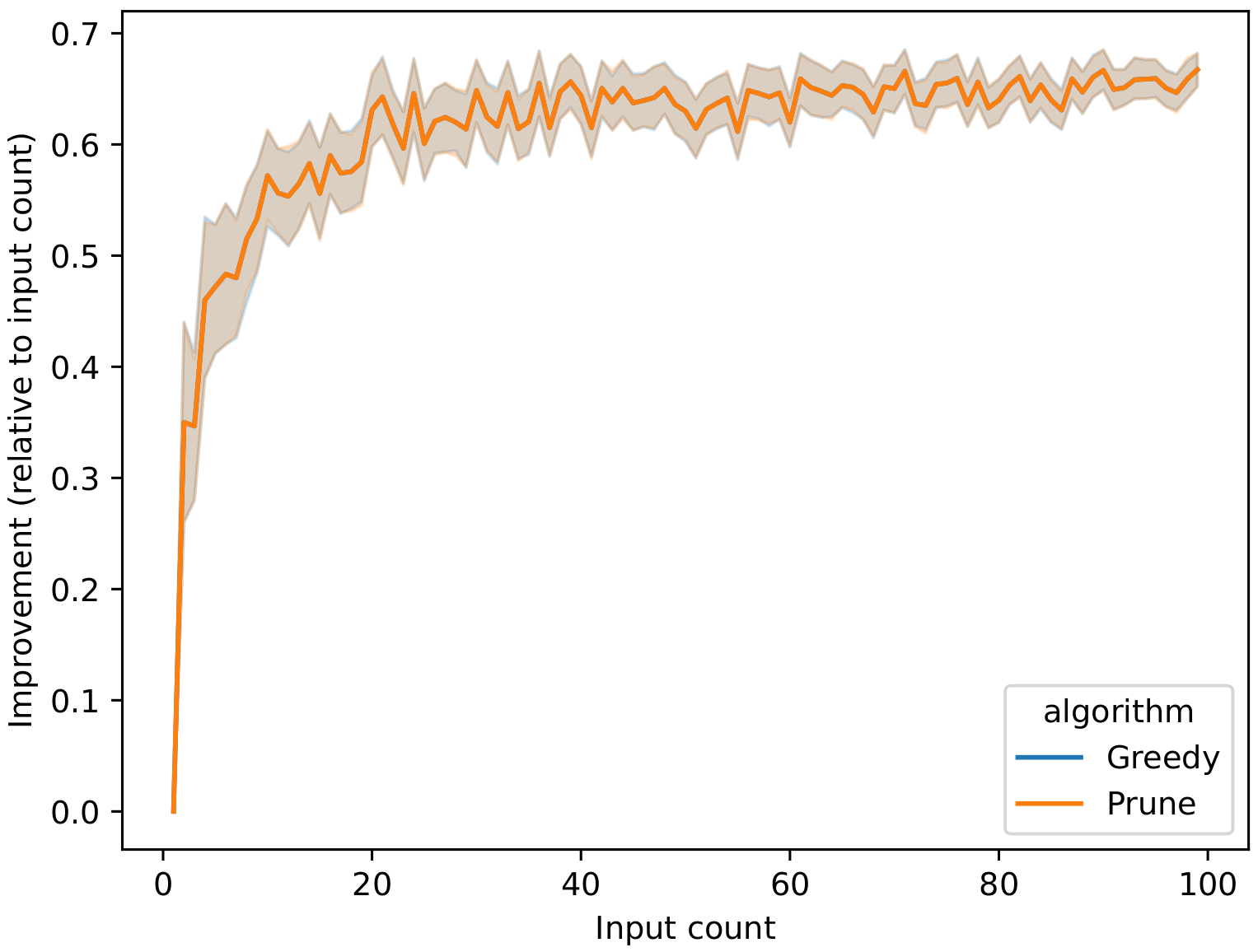}    
    \caption{}
    \end{subfigure}
\hfill
    \begin{subfigure}[b]{0.33\textwidth}
    \centering
    \includegraphics[width=\textwidth]{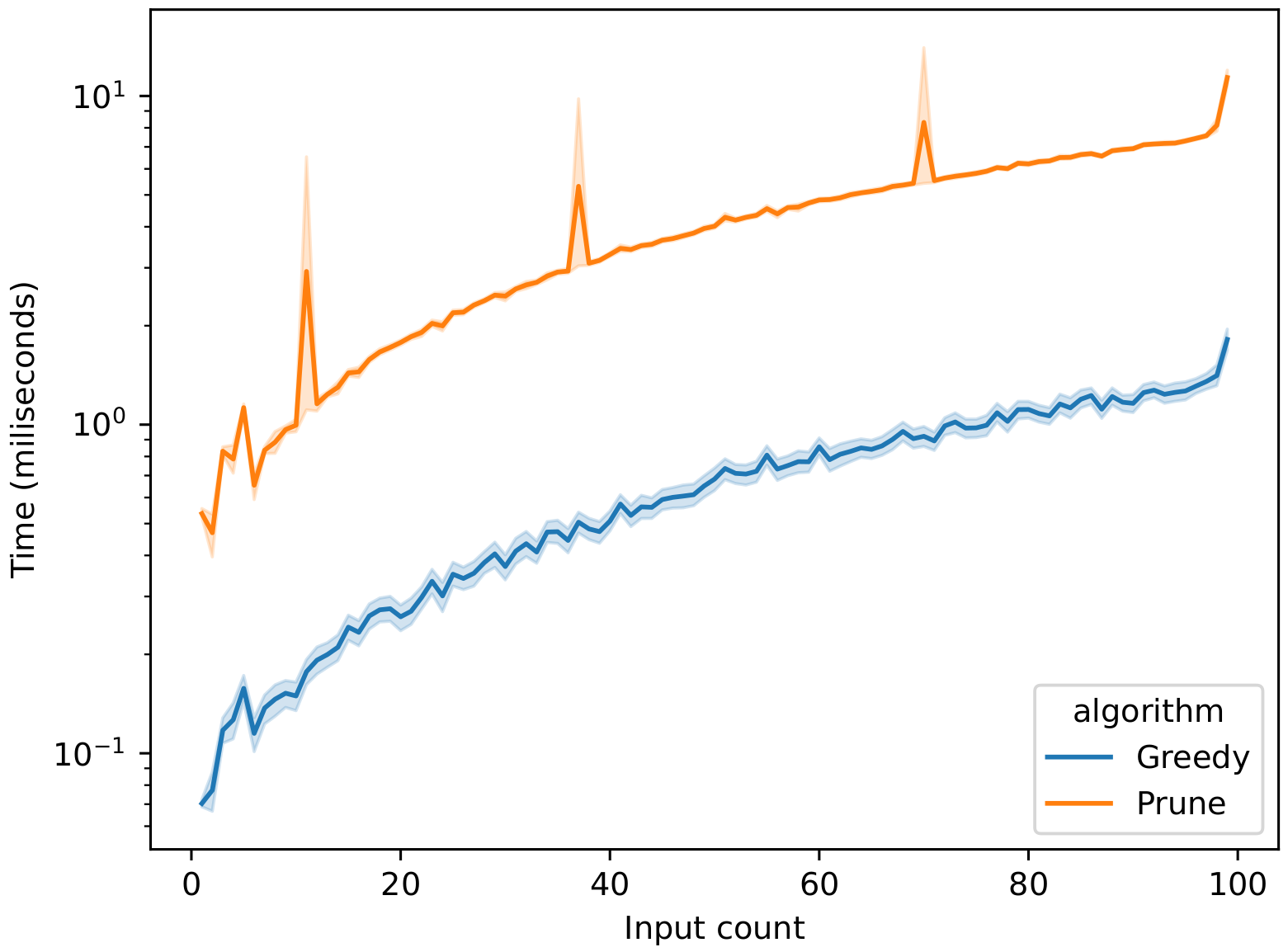}    
    \caption{}
    \end{subfigure}
\caption{Plots showing the (a) success rate, (b) improvement, and (c) time (in ms), for synthesis of 1-qubit unitaries with the Clifford+T gate set. The X-axis has the input T-count. The solid lines show the average metric for each data point (i.e. input T-count). The shaded region shows the variance around each point. ``Prune'' refers to the MIN-T-SYNTH algorithm in \cite{2021_MM}. ``Greedy'' refers to the approaches taken during the inference phase, as described in Section \ref{subsec:inferBenchmark}.   }
    \label{fig:t_plots123_1}
\end{figure}

We have generated 100 random unitaries per data point. Each data point corresponds to a particular input non-Clifford count for a specific universal gate set. We let our algorithms run for at most 60s for 1 and 2 qubit unitaries; and 180s for 3 and 4 qubit unitaries. Whenever our algorithm outputs a circuit within this time we refer to it as a "successful implementation". Now we compare these two counts and define the quality of our output circuits in terms of the improvement factor as defined below.
\begin{eqnarray}
    \text{Improvement} = 1 - \frac{\text{Output non-Clifford count}}{\text{Input non-Clifford count}}
    \label{eqn:improve}
\end{eqnarray}
We have also implemented the same input unitaries with previous algorithms \cite{2021_GRT, 2021_MM}. Out of these \cite{2021_GRT, 2021_MM} provably output the 2-qubit CS-count-optimal and 1-qubit T-count-optimal circuits, respectively. For unitaries on larger number of qubits we have considered the state-of-the-art  heuristic algorithm in \cite{2021_MM} (MIN-T-SYNTH) since it is faster than the provable ones and hence can be implemented within reasonable amount of time. For each universal gate set we have compared the performances of the algorithms using the following 3 metrics. 
\begin{enumerate}
    \item[(i)] \emph{Success Rate} : For each set of unitaries with a specific input non-Clifford-count (i.e. each data point), the success rate reflects the fraction of the unitaries that could be implemented within a specific time. Specifically,
    \begin{eqnarray}
        \text{Success Rate} = \frac{ \#\text{Successful implementations} }{ \text{Total } \#\text{unitaries} }.
        \label{eqn:succFactor}
    \end{eqnarray}

    \item[(ii)] \emph{Improvement} : This gives the reduction in output non-Clifford count compared to input non-Clifford count (Equation \ref{eqn:improve}), and hence can be regarded as a metric to evaluate the quality of the output circuit with respect to the non-Clifford count. 

    \item[(iii)] \emph{Time} taken to implement a unitary.  
\end{enumerate}
For both the metrics (ii) and (iii) only the successful implementations have been considered. Apart from the random unitaries, we have implemented some benchmark unitaries and compared the performance with previous known results.

\begin{figure}[htb]
    \begin{subfigure}[b]{0.3\textwidth}
    \centering
    \includegraphics[width=\textwidth]{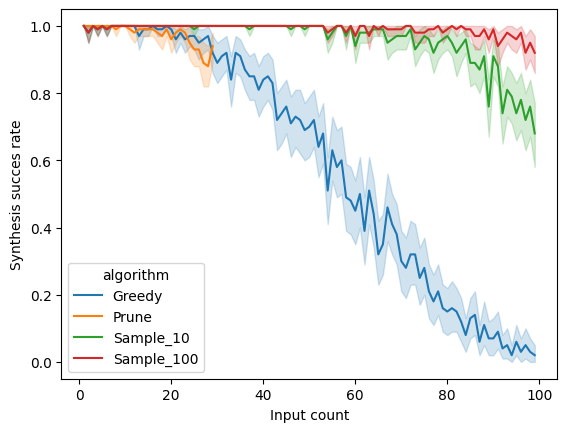}    
    \caption{}
    \end{subfigure}
\hfill
    \begin{subfigure}[b]{0.33\textwidth}
    \centering
    \includegraphics[width=\textwidth]{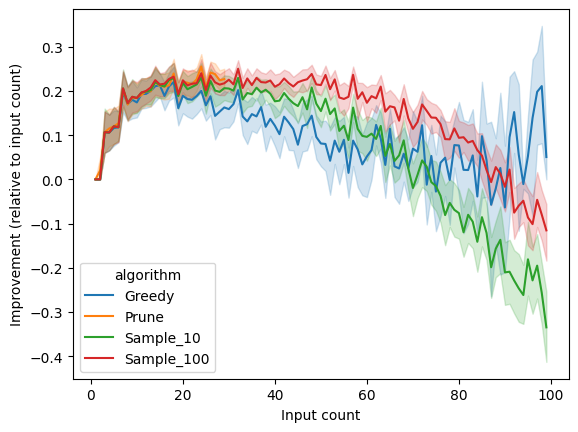}    
    \caption{}
    \end{subfigure}
\hfill
    \begin{subfigure}[b]{0.33\textwidth}
    \centering
    \includegraphics[width=\textwidth]{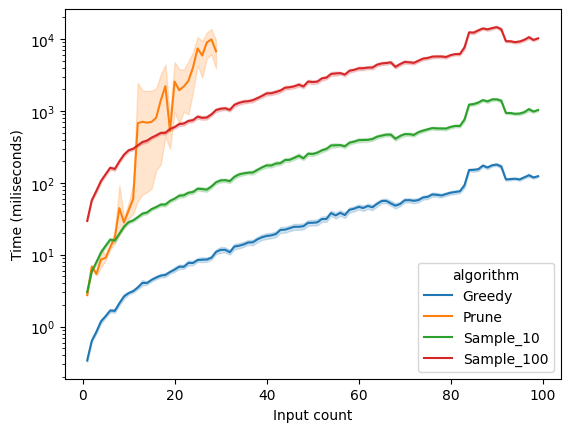}    
    \caption{}
    \end{subfigure}
\hfill
    \begin{subfigure}[b]{0.3\textwidth}
    \centering
    \includegraphics[width=\textwidth]{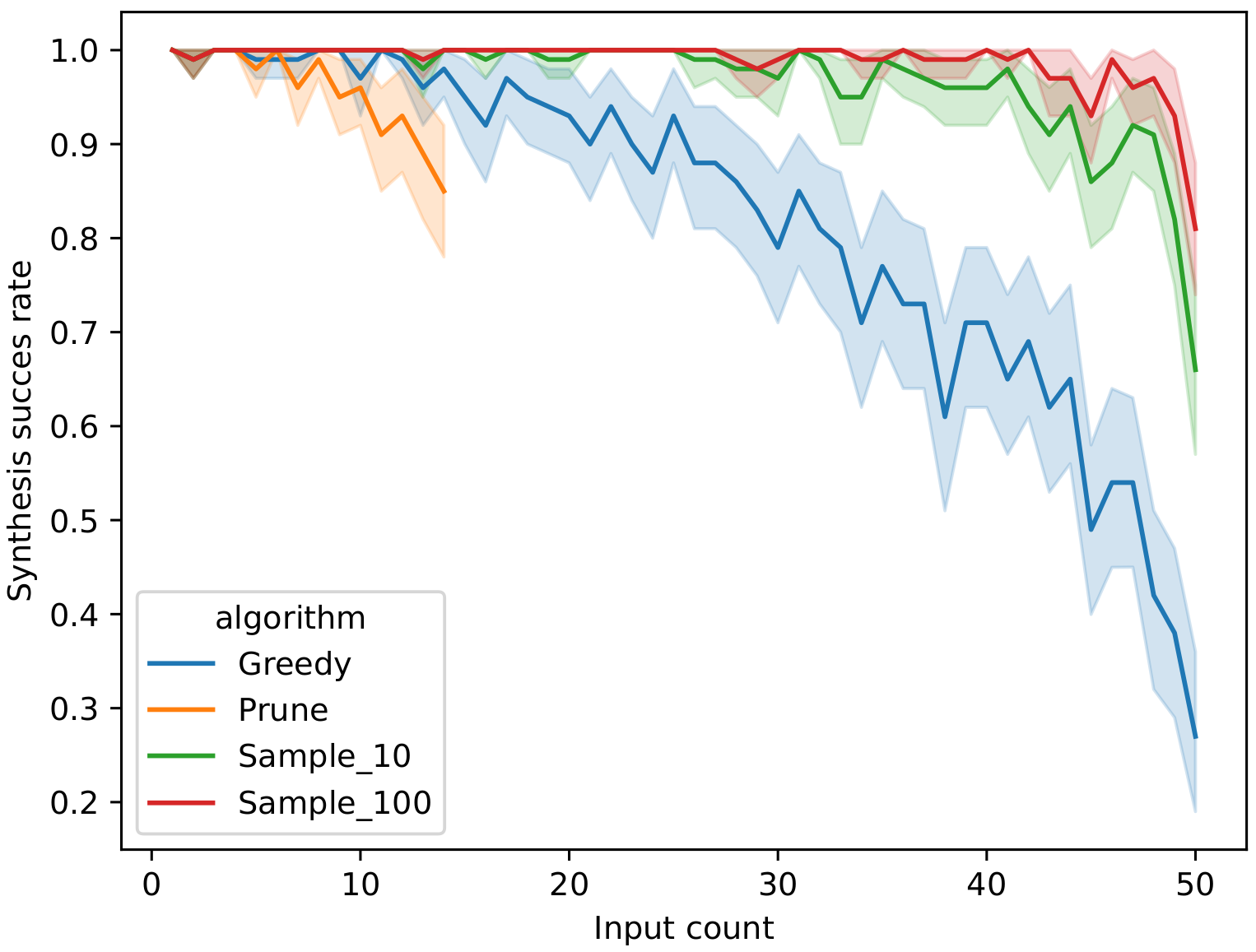}    
    \caption{}
    \end{subfigure}
\hfill
    \begin{subfigure}[b]{0.33\textwidth}
    \centering
    \includegraphics[width=\textwidth]{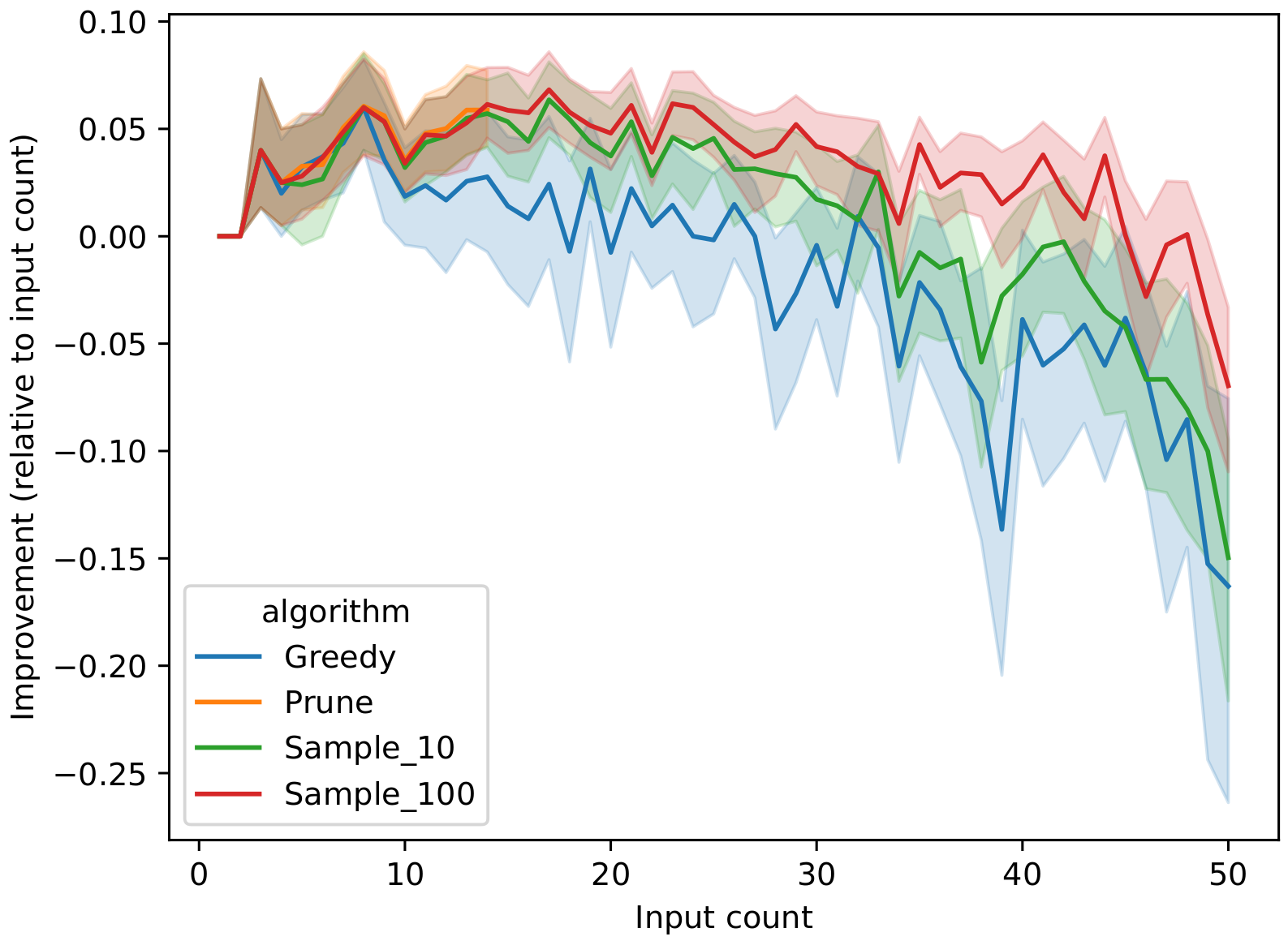}    
    \caption{}
    \end{subfigure}
\hfill
    \begin{subfigure}[b]{0.33\textwidth}
    \centering
    \includegraphics[width=\textwidth]{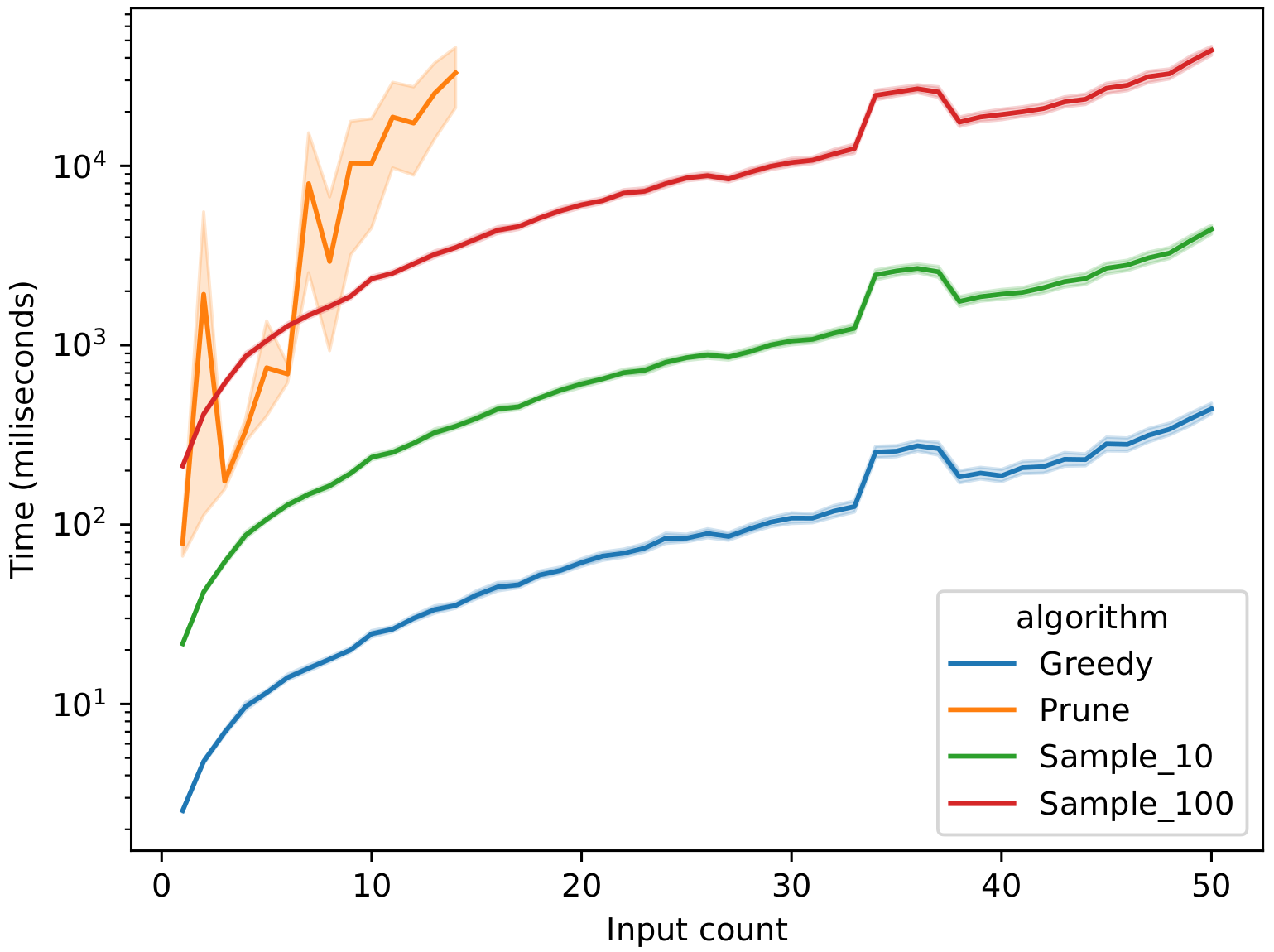}    
    \caption{}
    \end{subfigure}
    
    \caption{Plots showing the (i) success rate ((a),(d)), (ii) improvement ((b),(e)), and (iii) time in ms ((c),(f)), for  synthesis of 2 and 3-qubit unitaries with the Clifford+T gate set. The X-axis has the input T-count. Plots (a), (b), (c) are for 2-qubit unitaries, (d), (e), (f) are for 3-qubit unitaries. The solid lines show the average metric for each data point (i.e. input T-count). The shaded region shows the variance around each point. ``Prune" refers to the MIN-T-SYNTH algorithm in \cite{2021_MM}. ``Greedy", ``Sample$_{-}$10" and ``Sample$_{-}$100" refer to the approaches taken during the inference phase, as described in Section \ref{subsec:inferBenchmark}.   }
    \label{fig:t_plots123}
\end{figure}

\begin{figure}[h]
    \begin{subfigure}[b]{0.3\textwidth}
    \centering
    \includegraphics[width=\textwidth]{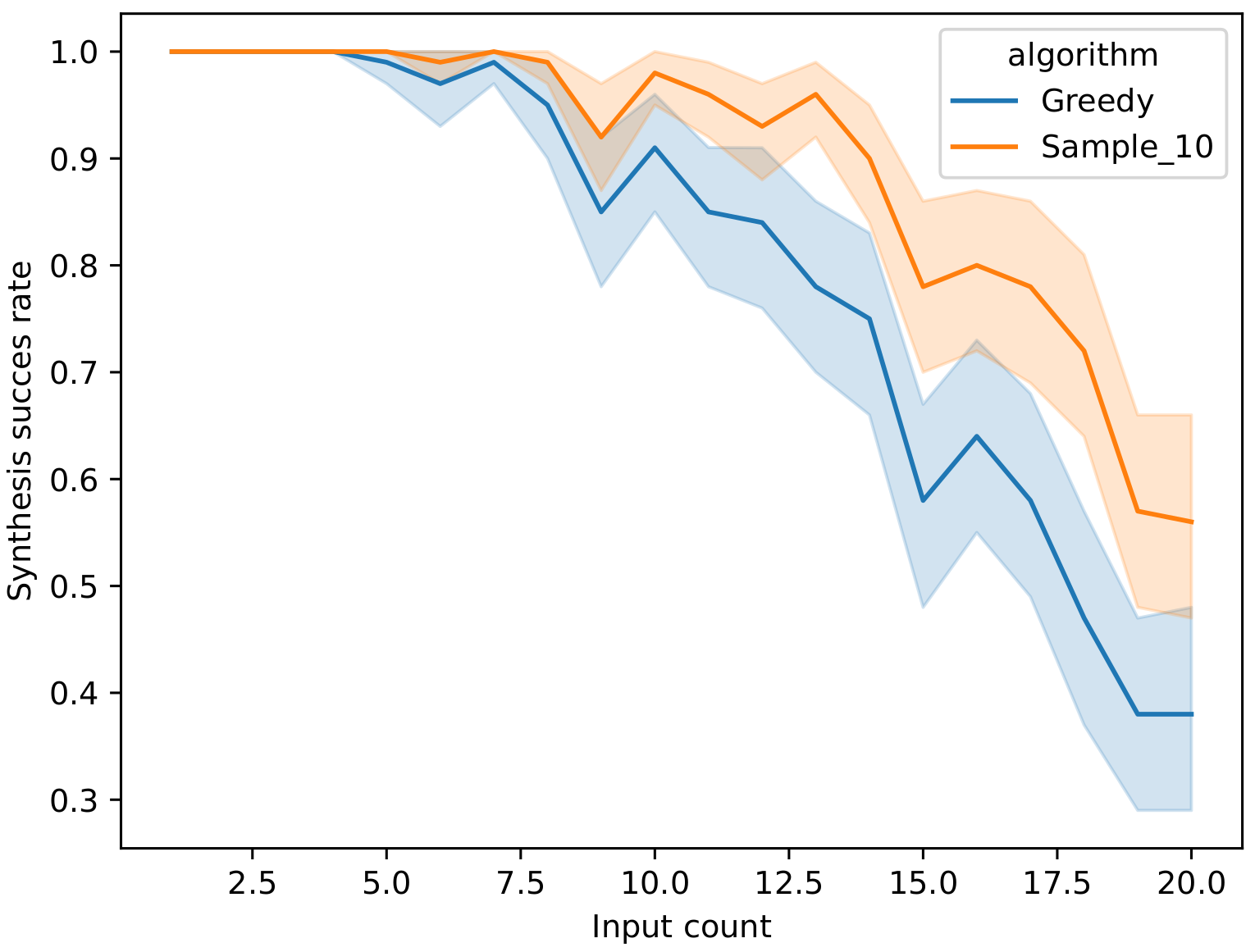}    
    \caption{}
    \end{subfigure}
\hfill
    \begin{subfigure}[b]{0.33\textwidth}
    \centering
    \includegraphics[width=\textwidth]{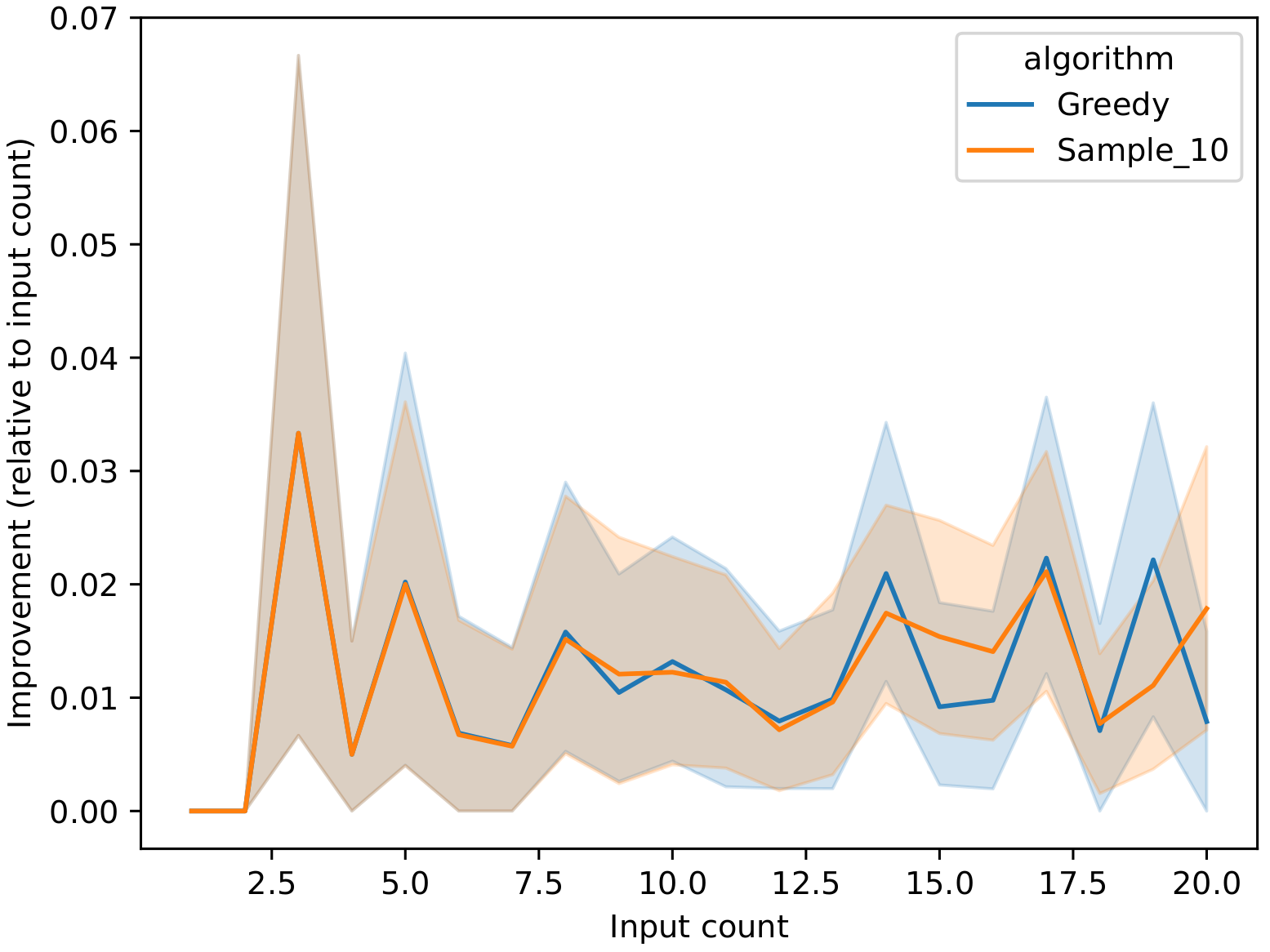}    
    \caption{}
    \end{subfigure}
\hfill
    \begin{subfigure}[b]{0.33\textwidth}
    \centering
    \includegraphics[width=\textwidth]{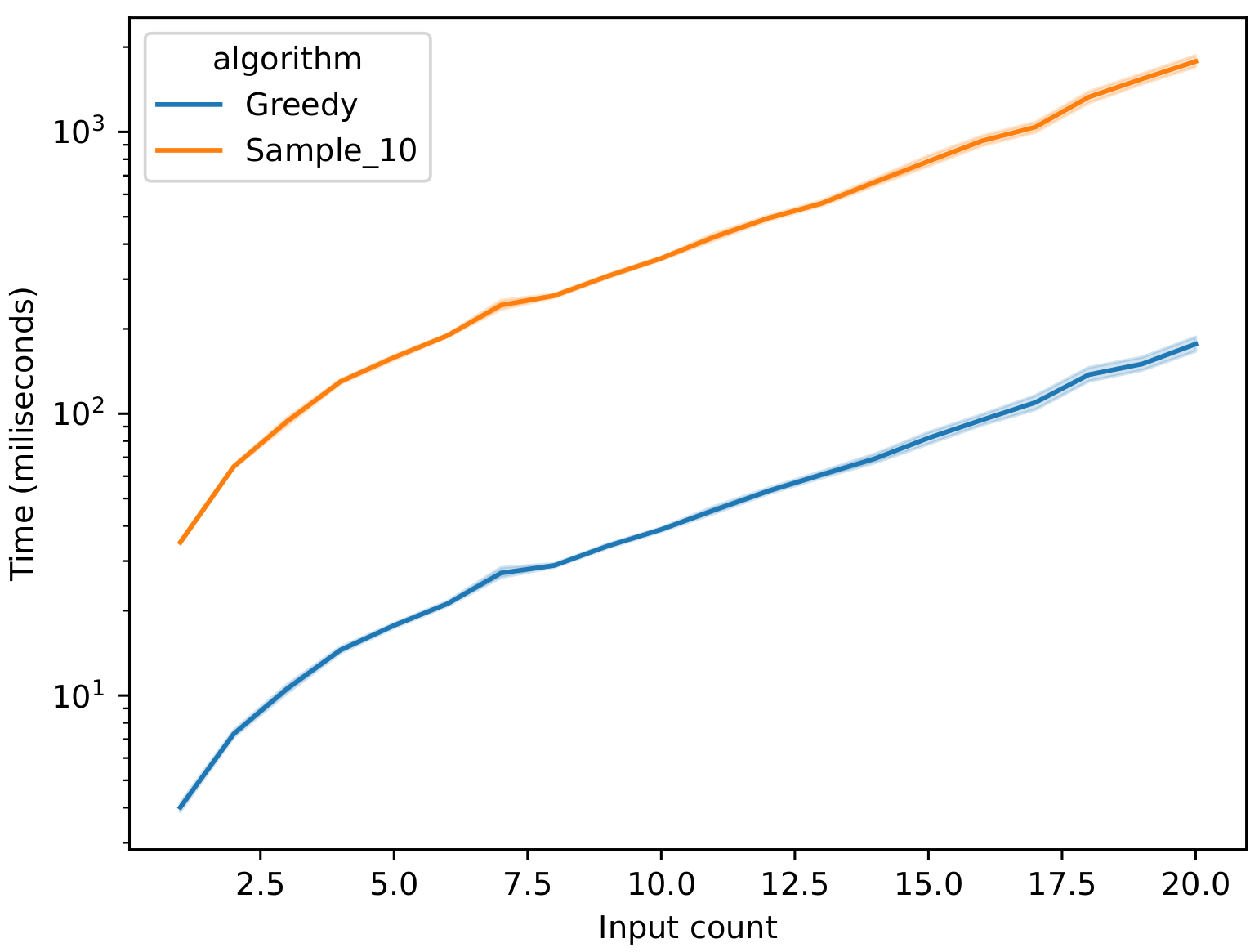}    
    \caption{}
    \end{subfigure}
    
    \caption{ Plots showing the (a) success rate, (b) improvement, and (c) time (in ms), for synthesis of 4-qubit unitaries with the Clifford+T gate set. The X-axis has the input T-count. The solid lines show the average metric for each data point (i.e. input T-count). The shaded region shows the variance around each point. "Greedy", "Sample$_{-}$10" and "Sample$_{-}$100" refer to the approaches taken during the inference phase, as described in Section \ref{subsec:inferBenchmark}. }
    \label{fig:t_plots45}
\end{figure}

\paragraph{T-count-optimal synthesis : } We have generated 1, 2, 3 and 4 qubit random unitaries with input T-count at most 100, 100, 50 and 20, respectively. We have synthesized these unitaries with our RL algorithm and MIN-T-SYNTH \cite{2021_MM}. The plots showing and comparing the performances of these algorithms have been given in Figures \ref{fig:t_plots123_1}, \ref{fig:t_plots123} and \ref{fig:t_plots45}. We have implemented three approaches - Greedy, Sample$_{-}$10 and Sample$_{-}$100 (refer Section \ref{subsec:inferBenchmark}). In the plots we refer to the algorithm MIN-T-SYNTH as Prune, due to space constraints.  We observe the following.

(a) For the 1-qubit case we achieve a success rate of 1, average improvement factor of about 0.7 and the time varies linearly with the T-count (Figure \ref{fig:t_plots123_1} (a)-(c)). We can say that we have obtained the T-count-optimal circuit for all unitaries because in all cases the $\sde_{\sqrt{2}}$ of the channel representation of the input unitary is equal to the output T-count \cite{2014_GKMR}. Additionally, MIN-T-SYNTH gives the same output T-count in all cases, as is also evident from the plots in Figure \ref{fig:t_plots123_1}(a), (b) and we know that the circuits returned by MIN-T-SYNTH for the 1-qubit unitaries are provably T-count-optimal. But our RL algorithm is about 10 times faster than MIN-T-SYNTH.  

(b) Our algorithm with 100 samples (Sample$_{-}$100) achieves a success rate of nearly 1 till about input T-count 90 for the 2-qubit unitaries, after which the success rate drops to about 0.93 at input T-count 100 (Figure \ref{fig:t_plots123}(a)). For 3-qubit unitaries the success rate is nearly 1 till input T-count 42, after which the average success rate hovers around 0.9 till input T-count about 49 (Figure \ref{fig:t_plots123}(d)). For 4-qubit unitaries the success rate is nearly 1 till input T-count 10 (Figure \ref{fig:t_plots45}(a)). In all the cases the success rate increases with the number of samples. This is much better than the success rate of MIN-T-SYNTH.

In all the cases the improvement factor is mostly non-zero, implying the output circuits have fewer T gates. Improvement factor of MIN-T-SYNTH is comparable till the limited range it succeeds (Figure \ref{fig:t_plots123}(b), \ref{fig:t_plots123}(e), \ref{fig:t_plots45}(b)).  We did not implement larger unitaries with MIN-T-SYNTH because it took more than 180s.

For the 2-qubit case the running time of MIN-T-SYNTH is less than Sample$_{-}$100 till input T-count about 15 (Figure \ref{fig:t_plots123}(c)). But then it quickly grows higher, while the latter has a (roughly) linear growth. For the 3-qubit case, MIN-T-SYNTH take more time, except for some few small values of input T-count (Figure \ref{fig:t_plots123}(f)). 

(c) Here we mention that for the successful implementations we have verified if the sequence of generating set unitaries output by our RL based algorithms is the same as the sequence ouptut by the heuristic algorithm MIN-T-SYNTH. We have noticed that in most cases they do. Such kind of studies throw some light on the existing conjectures proposed in \cite{2021_MM} and will help in further developing the existing ones or new ones for this and other problems. 

Now, let us consider the algorithm in \cite{2024_RDUetal}, which uses RL and to the best of our knowledge, this is the only prior ML algorithm that optimizes T-count while synthesizing circuits for given unitaries. From the descriptions and data given in the paper we observe the following.
\begin{enumerate}
    \item[(a)] \cite{2024_RDUetal} implements 2, 3 and 4-qubit unitaries with T-count at most 20 and 5-qubit unitaries with T-count at most 15. Thus we are able to implement much larger unitaries with higher T-count. 

    \item[(b)] The success factor in \cite{2024_RDUetal} is defined in terms of a quantity that depends on the overlap of the output unitary with the input unitary. When this overlap is high enough the implementation is deemed successful. So there is a provision for the experiment to be considered successful even if the output unitary is not equal to the target unitary (up to a global phase). Though, the authors do state that all the unitaries considered by them have been exactly implemented. Here we note that the target unitaries in \cite{2024_RDUetal} are all exactly implementable since they have been generated from randomly sampled circuits.
    
    Further, in \cite{2024_RDUetal} there is a time-out condition of 400s, that is higher than our 180s. From the plots given in Figure 4 of \cite{2024_RDUetal} we notice that for the 2 and 3-qubit cases the success probability is nearly 1 till T-count 10, but drops to about 0.5 at T-count 20. For 4-qubit unitaries the success probability is 1 till T-count 8 and drops relatively more quickly to about 0.15 at T-count 20. 

    Now, if we compare this with our success ratio/probability, as depicted in Figures \ref{fig:t_plots123}(a), \ref{fig:t_plots123}(d) and \ref{fig:t_plots45}(a), we infer that our algorithm performs significantly better, although we have a much lower time-out condition.

    \item[(c)] Figure 5 in \cite{2024_RDUetal} shows that unitaries with at most 57 gates have been synthesized. We can synthesize much larger circuits because we use the generating set formalism. As explained earlier, given a target unitary $U$, we first implement a circuit for $U$ (modulo a trailing Clifford) with unitaries from $\gen_T$ (Equation \ref{eqn:genT}). Later each $\gen_T$ unitary and the trailing Clifford is implemented with the Clifford+T gate set. This compression-expansion procedure enables us to implement much larger circuits. If we assume that on an average the Clifford+T gate count of each $R(P)$ is $O(n)$, then the average total gate count of unitaries with T-count at most $t$ is at least $O(nt)$. This is significantly much larger than the gate-counts of circuits implemented by $\cite{2024_RDUetal}$ or other papers. For example, let us consider the unitary compilation results in \cite{2024_FMB}. The largest circuit synthesized has 3 qubits, 12 gates and is over the gate set $\{\had, \CNOT, \Z, \X, \tof, \swap\}$. This is significantly less than what we achieve.
\end{enumerate}

\begin{figure}[h]
    \begin{subfigure}[b]{0.3\textwidth}
    \centering
    \includegraphics[width=\textwidth]{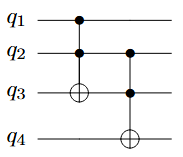}    
    \caption{$U_1$}
    \end{subfigure}
\hfill
    \begin{subfigure}[b]{0.33\textwidth}
    \centering
    \includegraphics[width=\textwidth]{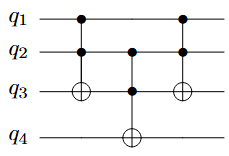}    
    \caption{$U_2$}
    \end{subfigure}
\hfill
    \begin{subfigure}[b]{0.33\textwidth}
    \centering
    \includegraphics[width=\textwidth]{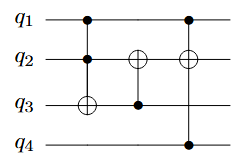}    
    \caption{$U_3$}
    \end{subfigure}
\hfill
    \begin{subfigure}[b]{0.33\textwidth}
    \centering
    \includegraphics[width=\textwidth]{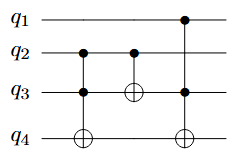}    
    \caption{$U_4$}
    \end{subfigure}
\hfill
    \begin{subfigure}[b]{0.3\textwidth}
    \centering
    \includegraphics[width=\textwidth]{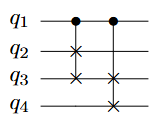}    
    \caption{$U_5$}
    \end{subfigure}
    
    \caption{Circuits for the benchmark unitaries defined in Equation \ref{eqn:benchmark} }
    \label{fig:benchmark}
\end{figure}

\paragraph{Other benchmark unitaries :} We have also implemented some benchmark unitaries obtained from previous papers, as well as some new unitaries, as listed below. $\tof_{(q_i,q_j;q_k)}$ implies a Toffoli with controls on qubits $q_i, q_j$ and target on $q_k$. $\fred_{(q_i;q_j,q_k)}$ implies a Fredkin or CSWAP (controlled-SWAP) with control on qubit $q_i$ and SWAP between qubits $q_j$ and $q_k$. The circuits have been shown in Figure \ref{fig:benchmark}. 
\begin{eqnarray}
    U_1 &=& \tof_{(q_1,q_2;q_3)}\cdot\tof_{(q_2,q_3;q_4)}  \nonumber \\
    U_2 &=& \tof_{(q_1,q_2;q_3)}\cdot\tof_{(q_2,q_3;q_4)}\cdot\tof_{(q_1,q_2;q_3)}  \nonumber \\
    U_3 &=& \tof_{(q_1,q_2;q_3)}\cdot\CNOT_{(q_3;q_2)}\cdot\tof_{(q_1,q_4;q_2)}   \nonumber \\
    U_4 &=& \tof_{(q_2,q_3;q_4)}\cdot\CNOT_{(q_2;q_3)}\cdot\tof_{(q_1,q_3;q_4)}   \nonumber \\
    U_5 &=& \fred_{(q_1;q_2,q_3)}\cdot\fred_{(q_1;q_3,q_4)}
    \label{eqn:benchmark}
\end{eqnarray}
The results have been provided in Table \ref{tab:benchmark}. We have compared the running time and T-count with those quoted in \cite{2021_MM} and \cite{2024_RDUetal} (for unitaries implemented by them). Our running time is much better than both these algorithms and for $U_1$ we get a lower T-count than \cite{2024_RDUetal}. From previous papers we can say that we obtain the optimal T-count for the unitaries tabulated till $U_2$. For the remaining, we note that the obtained T-count is much less than the number of T-gates obtained by plugging in the T-count-optimal decomposition of Toffoli and Fredkin in these unitaries.

We note that, to the best of our knowledge, we are not aware of any resynthesis algorithms that can give lower T-count than what has been obtained by us. In fact, \cite{2014_AMM} gives a T-count of 12 for $U_1$, which is higher than our result. We note that the circuits of the above benchmark unitaries include H gate and so T-count optimizing resynthesis algorithms like TODD \cite{2018_HT} or its later variants \cite{2024_RLBetal, 2025_V} would have introduced ancilla for gadgetization. Our constructions do not introduce any extra ancilla. 

\begin{table}[h]
    \centering
    \footnotesize
    \begin{tabular}{|p{1.5cm}|p{1.5cm}|p{1.5cm}|p{1.5cm}|p{1.5cm}|p{1.5cm}|p{1.5cm}|p{1.5cm}|}
    \hline
       \textbf{Unitaries}  & \textbf{$\#$Qubits}  & \textbf{T-count (Our algo)} & \textbf{T-count (\cite{2021_MM})} & \textbf{Tcount (\cite{2024_RDUetal})} & \textbf{Time (Our algo)} & \textbf{Time (\cite{2021_MM}}) & \textbf{Time (\cite{2024_RDUetal})}\\
       \hline
        Toffoli & 3 & 7 & 7 & 7 & 3.862s & 5.75s & 25.92s\\
        \hline
        Fredkin & 3 & 7 & 7 & 7 & 3.784s & 5.9s & 16.96s \\
        \hline 
        Peres & 3 & 7 & 7 & 7 & 3.827 & 5.74s & 16.87s \\
        \hline
        Quantum OR & 3 & 7 & 7 & 7 & 3.828s & 5.74s & 16.56s \\
        \hline
        Negated Toffoli & 3 & 7 & 7 & 7 & 3.483s & 5.75s & 16.78s \\
        \hline
        1-Bit adder & 4 & 7 & 7 & 7 & 12.449s & 429.17s & 18.53s \\
        \hline
        $U_1$ & 4 & 11 & 11 & 12 & 11.677s & 2.17h & 19.27s \\
        \hline
        $U_2$ & 4 & 7 & 7 & 7 & 11.677s & 391.27s & 31.50s \\
        \hline
        $U_3$ & 4 & 11 & - & - & 11.677s & - & - \\
        \hline
        $U_4$ & 4 & 7 & - & - & 11.677s & - & - \\
        \hline
        $U_5$ & 4 & 8 & - & - & 3.3m & - & - \\
        \hline
    \end{tabular}
    \caption{T-count and time required to implement some benchmark unitaries  with our RL based algorithm, MIN-T-SYNTH \cite{2021_MM} and the previous RL based algorithm in \cite{2024_RDUetal}. Unitaries $U_1$-$U_5$ have been defined in Equation \ref{eqn:benchmark}. A "-" indicates unavailability of data from previous papers. }
    \label{tab:benchmark}
\end{table}

\begin{figure}[h]
    \begin{subfigure}[b]{0.3\textwidth}
    \centering
    \includegraphics[width=\textwidth]{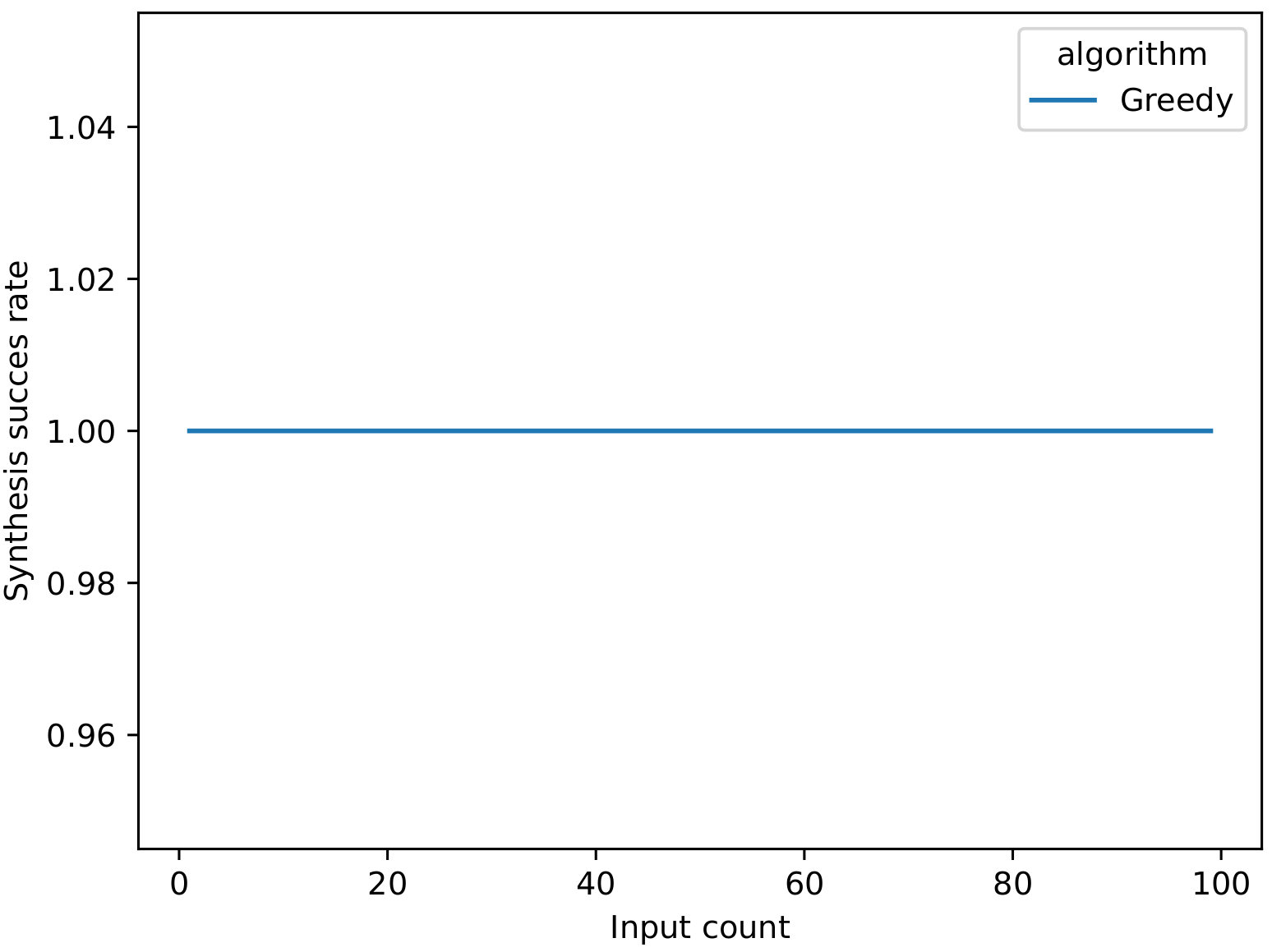}    
    \caption{}
    \end{subfigure}
\hfill
    \begin{subfigure}[b]{0.33\textwidth}
    \centering
    \includegraphics[width=\textwidth]{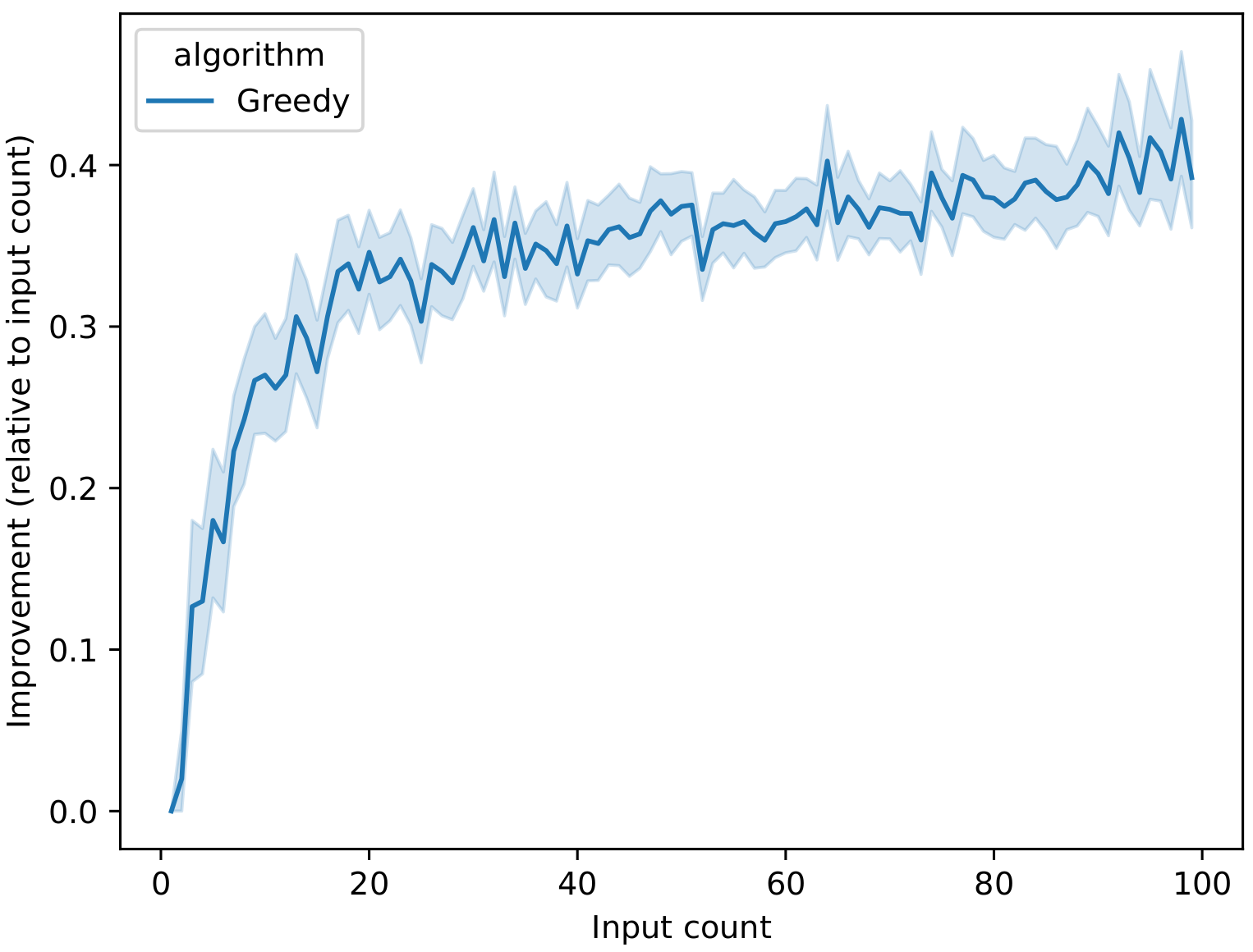}    
    \caption{}
    \end{subfigure}
\hfill
    \begin{subfigure}[b]{0.33\textwidth}
    \centering
    \includegraphics[width=\textwidth]{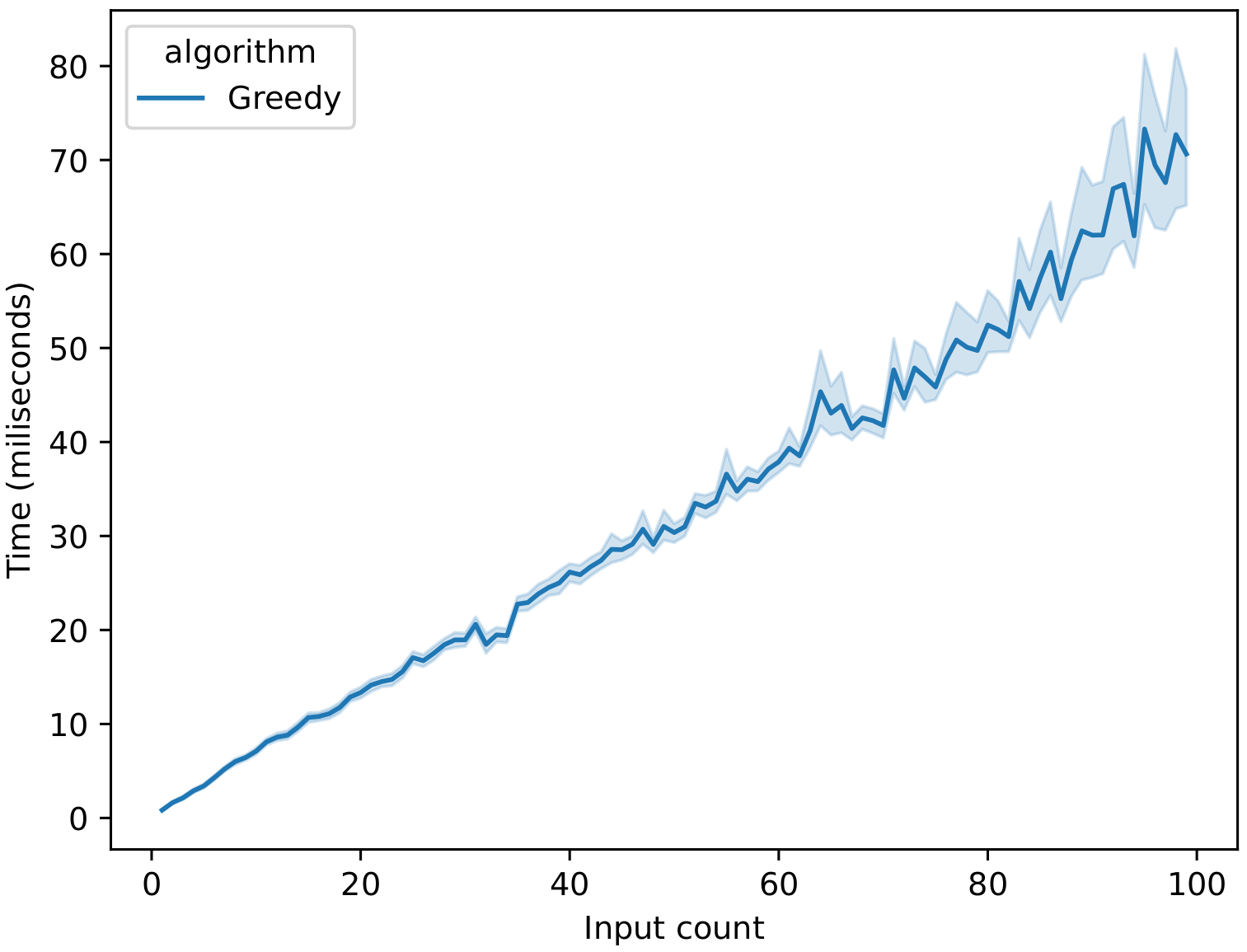}    
    \caption{}
    \end{subfigure}
    
    \caption{Plots showing the (a) success rate, (b) improvement, and (c) time (in ms), for synthesis of 2-qubit unitaries with the Clifford+CS gate set. The X-axis has the input CS-count. The solid blue line shows the average metric for each data point (i.e. input CS-count). The shaded region shows the variance around each point. "Greedy" refers to the greedy approach taken during the inference phase, as described in Section \ref{subsec:inferBenchmark}. }
    \label{fig:cs_plots}
\end{figure}

\paragraph{CS-count-optimal synthesis : } We have synthesized quantum circuits for 2-qubit unitaries with CS-count at most 100. Plots of the metrics reflecting the performance of our algorithm have been given in Figure \ref{fig:cs_plots}. We have observed the following.
\begin{enumerate}
    \item[(a)] For 2-qubit unitaries the success factor is 1, the improvement factor is strictly positive, implying we are able to synthesize circuits for all target unitaries and the CS-count of the output circuits is at most the CS-count of the input circuits. In fact, from the plot in Figure \ref{fig:cs_plots}(b) we see that the CS-count actually improves for most unitaries. 

    \item[(b)] We achieve a linear time complexity. Similar time complexity for the specific case of 2-qubit unitaries, has been accomplished in \cite{2021_GRT}, where unitaries in $SU(4)$ have been mapped to elements in $SO(6)$ by utilizing the exceptional isomorphism $SU(4)\cong Spin(6)$. But previous algorithms working with channel representation, for example, \cite{2024_Mcs}, do not exhibit such efficiency for the 2-qubit case.
\end{enumerate}

\subsection{Application in reducing the asymptotic complexity of arithmetic circuits}

In this section we illustrate how optimized circuits of smaller unitaries can reduce the asymptotic gate complexity of larger circuits for specific tasks. This in turn can reduce the asymptotic gate complexity of many quantum algorithms where these operations appear as sub-routines. 

\begin{figure}[h]
    \begin{subfigure}[b]{0.35\textwidth}
    \centering
    \includegraphics[width=\textwidth]{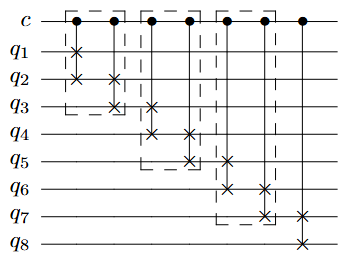}   
    \caption{}
    \end{subfigure}
\hfill
    \begin{subfigure}[b]{0.6\textwidth}
    \centering
    \includegraphics[width=\textwidth]{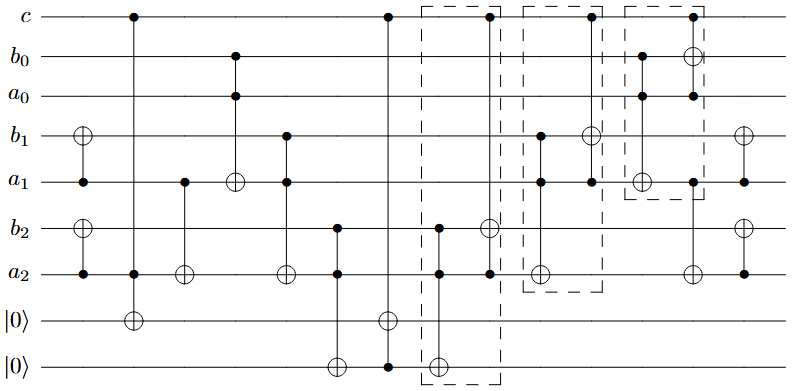}    
    \caption{}
    \end{subfigure}
    
    \caption{(a) A quantum circuit for controlled cyclic shift of basis vectors. (b) A quantum circuit for controlled addition. }
    \label{fig:app}
\end{figure}

\paragraph{Controlled cyclic shift : } In many applications, we require to perform a cyclic shift of basis states. A quantum circuit for this purpose has been given in \cite{2018_HSRS}. We have re-drawn this in Figure \ref{fig:app}(a). Given a particular basis state, this circuit shifts the bit positions by $s$ places. For example, consider a $n+1$-qubit circuit, whose control is qubit $c$ and the other qubits $q_1,\ldots, q_n$ are in a computational basis state $\bigotimes_{j=1}^n\ket{a_j}$, where $a_j\in\{0,1\}$. After the operation of the Fredkin or CSWAP gates, the qubits are in state 
\begin{eqnarray}
\bigotimes_{j=1}^n\ket{b_j},\qquad \text{where } b_j = a_{j+1},\quad \text{for } j = 1,\ldots,n-1 \quad\text{ and } b_n = a_1; \nonumber
\end{eqnarray}
if and only if $\ket{c} = \ket{1}$. If the qubits are in superposition of basis states, then similar cyclic shift occurs for each basis state. We require $n-1$ Fredkin gates in this circuit. If we plug in the T-count-optimal circuit of a Fredkin gate, then we obtain a circuit with $7(n-1)$ T gates (Table \ref{tab:benchmark}). Instead, if we plug in the T-count-optimized circuit of $U_5$ (Equation \ref{eqn:benchmark} and Figure \ref{fig:benchmark}(e)) (boxed in Figure \ref{fig:app}(a)), then we obtain a circuit with T-count 
\begin{eqnarray}
    4(n-1), &\quad& \text{if }n\text{ is odd};  \nonumber\\
  4(n-2)+7 = 4n-1, &\quad& \text{if } n\text{ is even}.  \nonumber
\end{eqnarray}
This ensures a reduction in the T-count by about $43\%$ of the initial count.   

\paragraph{Controlled addition :} Controlled addition appears in many quantum algorithms and a popular quantum circuit for it can be found in Figure 5 of \cite{2018_MT}. The purpose of this circuit is to add two $n$-bit integers - $a = (a_{n-1},\ldots, a_0)$ and $b = (b_{n-1},\ldots, b_0)$, conditioned on the value of a control bit, $c$. $\ket{a} = \ket{a_{n-1},\ldots,a_0} = \bigotimes_{i=0}^{n-1}\ket{a_i}$, $\ket{b} = \ket{b_{n-1},\ldots,b_0} = \bigotimes_{i=0}^{n-1}\ket{b_i}$ and $\ket{c}$ are input states. There are two additional ancilla, initialized to $\ket{0}$. The circuit operates as follows.
\begin{eqnarray}
    \ket{c}\ket{0}\ket{0}\ket{b}\ket{a}&\mapsto&\ket{c}\ket{0}\ket{cs_n\oplus\conj{c} 0}\ket{c s_{n-1}\oplus\conj{c}b_{n-1},\ldots,cs_0\oplus\conj{c}b_0}\ket{a}\qquad[\text{if }c=1];   \nonumber \\
    \ket{c}\ket{0}\ket{0}\ket{b}\ket{a}&\mapsto&\ket{c}\ket{0}\ket{\conj{c}s_n\oplus c 0}\ket{\conj{c} s_{n-1}\oplus cb_{n-1},\ldots,\conj{c}s_0\oplus cb_0}\ket{a}\qquad[\text{if }c=0];   \nonumber 
\end{eqnarray}
For convenience, we have drawn a controlled-adder for 3-bit numbers in Figure \ref{fig:app}(b). For the controlled addition of $n$-bit numbers we require $(3n+2)$ Toffoli gates, and hence $(21n+14)$ T-gates \cite{2018_MT}. We observe that out of these, there are $n$ pairs of Toffoli gates such that each pair is of the form $U_1$ (Equation \ref{eqn:benchmark} and Figure \ref{fig:benchmark}(a)). We have boxed them in Figure \ref{fig:app}(b).If we plug in the circuits for $U_1$, then we obtain a T-count (Table \ref{tab:benchmark}) estimate of
\begin{eqnarray}
    7(n+2)+11n = 18n+14.    \label{eqn:T_ctrl_add}
\end{eqnarray}
 Hence we obtain a reduction of about $14.3\%$ of the initial T-count estimate. 

\paragraph{Multiplication :} Multiplication is an important arithmetic primitive that is an integral part of many quantum algorithms. We consider the quantum circuit for multiplying two $n$-bit integers given in Figure 7 of \cite{2018_MT}. It consists of $(n-1)$ controlled adders and an additional 7 Toffoli gates. The T-count estimate given in \cite{2018_MT} is $(21n^2-14)$. If we use the T-count estimate of controlled adder, as given in Equation \ref{eqn:T_ctrl_add}, then we obtain a T-count of
\begin{eqnarray}
    (18n+14)(n-1)+7 = 18n^2+3n-14.  \nonumber
\end{eqnarray}
Hence we obtain a reduction of $(3n^3-3n)$, that is about $14\%$ of the initial estimate, on an average.

Again, we note that we do not introduce any extra ancilla for any of the above circuits, as opposed to other T-count optimizing resynthesis algorithms \cite{2014_AMM, 2018_HT, 2024_RLBetal, 2025_V}.

 \section{Method}
\label{sec:method}

We target exact Clifford+T and Clifford+CS synthesis for small-$n$ qubit unitaries by casting the task as a single-player reinforcement learning (RL) problem. In this section we describe how we formulate the problem as a reinforcement learning problem, how we perform the training, and how we do the inference and benchmarking on a trained model.

At a high level, we follow the training framework described in \cite{2024_KVPetal}. An RL agent is given a target operator and selects a gate from a pre-defined gateset, which is then used to evolve the operator. This new operator is then provided to the RL agent, that chooses another gate, and this step by step process continues until an identity operator is reached, implementing the input operator as a circuit. The RL agent is trained by interacting with an RL environment that implements these state transitions, providing rewards when the identity is reached and penalties for each gate incurred. The training also proceeds by starting with easy instances and progressively increases difficulty as the model's success rate reaches a given threshold.

We introduce a few crucial additions and modifications to this framework that make the procedure work for the Clifford+T and Clifford+CS cases. These are listed below.
\begin{itemize}
    \item {\bf Channel representation.} Instead of using the unitary matrix representation (as in other works, for example, \cite{2024_FMB, 2024_KVPetal, 2024_RDUetal}), we describe the operators in the channel representation. This allows us to work with integers for exactly implementable unitaries (Section \ref{subsec:chanRep}), while previous papers work with complex numbers. This effectively discretizes the problem, and makes it easier for the RL model to distinguish between close but different states that may have very different circuit implementations.

    \item  {\bf Canonical observations.}
    The channel representation also allows us to canonicalize the operator by lexicographically sorting the columns, as described in the following definition of \emph{coset label}. 

    \begin{definition}[\textbf{Coset label}]
    Let $W\in\clifft_n^T$. The coset label of $W$ is a matrix, $\chan{W}^{(c)}$, that is obtained from $\chan{W}$ using the following procedure. (i) Rewrite $\chan{W}$ so that each non-zero entry has a common denominator $\sqrt{2}^k$, where $k = \sde_{\sqrt{2}}\left(\chan{W}\right)$. (ii) Modify each column of $\chan{W}$ as follows. Look at the first non-zero entry (from top to bottom), $v$, and express it in the form $v=\frac{a+b\sqrt{2}}{\sqrt{2}^k}$. If $a < 0$, or if $a = 0$ and $b < 0$, multiply every element of the column by $-1$. Else, keep the column unchanged. (iii) Permute the columns so that they are ordered lexicographically from left to right.
    \label{defn:cosetLabel}
\end{definition}

\begin{fact}[\cite{2014_GKMR}]
Let $W, V\in\clifft_n$. Then $\chan{W}^{(c)} = \chan{V}^{(c)}$ if and only if $W = VC$ for some $C\in\cliff_n$.
\label{fact:cosetLabel}
\end{fact}

Thus any two operators that differ only by a Clifford are mapped to the same representation. So the model only needs to ``recognize'' a single variant of a given unitary across all Clifford variations, greatly reducing the number of distinct states at each height of the tree.

    \item {\bf Generating set as actions.} Since we want to reduce the number of non-Clifford gates, we use the generating sets as the gate set instead of the usual Clifford+T or Clifford+CS (Section \ref{subsec:genSet}). This ensures that each step has the same cost (as each step contains one non-Clifford gate each), and reduces the depth of the search tree (since the Clifford operations are ``included'' in each step).

    \item {\bf Action masking from pruning heuristics.} We also introduce action masking based on previous work on pruning heuristics \cite{2021_MM, 2024_Mcs}. This allows us to discard actions that are known to be non-optimal gate choices at each step, effectively reducing the action space. We use the Divide-and-Select method A in \cite{2021_MM} (Clifford+T) and \cite{2024_Mcs} (Clifford+CS). By this method, the intermediate unitaries at any level of the tree are divided into two groups - one whose sde increases with respect to the parent unitary and the remaining. The set with the minimum cardinality are selected for expansion in the next level. The remaining are discarded. Further, in order to increase the efficiency of the Divide-and-Select procedure we have described a new algorithm in Section \ref{subsec:newDivSel}. 

    \item {\bf AlphaZero training with curriculum learning.} We combine single-player AlphaZero with curriculum learning by defining a reward function that is bounded but scales with the difficulty of the input, so the model can learn an estimate to the absolute distance of a given unitary to a Clifford.
\end{itemize}

In the following subsections we describe these methods in detail.

\subsection{Problem formulation}

Here we describe the dynamics of the RL environment and the RL neural network architecture.

\subsubsection{State representation}

The environment state is the channel representation of an $n$-qubit unitary $U$, denoted by $\chan{U}$, that is an array of shape $4^n\times 4^n$.
For Clifford+T implementation, the entries of $\chan{U}$ are integer triples $(a,b,k)$, as described in Section \ref{subsec:chanRep}. 
The objective is to reach the channel representation of a Clifford, a unitary described in Fact \ref{fact:chanRepCliff}, while minimizing the number of $R(P)$ unitaries (Equation \ref{eqn:genT}). This realizes a Clifford+T decomposition of the input unitary $U$, with minimal or near-minimal T-count. 

For Clifford+CS implementation, we use an equivalent representation but in this case each entry is described with just two integers $(a,k)$, as described in Section \ref{subsec:chanRep}. In the remainder of this section we describe the method for Clifford+T. The Clifford+CS variant is similar for the RL method with the difference of having one less integer variable in the representation and the use of unitaries $G_{P_1P_2}$ (Equation \ref{eqn:genCS}), instead of $R(P)$.

\subsubsection{Observation encoding}

Let $B$ denote the fixed bit-width for signed integers. For each entry, we concatenate the $B$-bit representations of $a$, $b$, and $k$ into a length-$3B$ binary vector. Stacking over all entries yields a tensor of shape $[4^n,\,4^n,\,3B]$.

Because synthesis is defined up to a Clifford, any column permutation corresponds to an equivalent state. To reduce input variability and promote invariance, we sort columns lexicographically (see Definition \ref{defn:cosetLabel}). To expose right- and left-multiplication symmetries, we also include the inverse of the channel: we form a second tensor by first transposing $X$, applying the same lexicographic column sorting, and re-encoding as above, then we stack this channel-wise with the original view. The final observation is
\begin{equation}
  \mathcal{O}\in\{0,1\}^{4^n\times 4^n\times 2\cdot 3B}.
\end{equation}

\subsubsection{Action space and transitions}

The action set is the full set of $n$-qubit Pauli strings $\mathcal{P}_n=\{I,X,Y,Z\}^{\otimes n}$ (size $4^n$) where each can be applied either from the left or from the right. Each action specifies the application of a $R(P)$ unitary, together with either $X\leftarrow R(P)X$ (\emph{left}) or $X\leftarrow XR^{\dagger}(P)$ (\emph{right}), which is equivalent (in the channel picture) to transposing, left-multiplying, and transposing back. The discrete action space thus has cardinality $2\cdot 4^n$. After each step, we update the $(a,b,k)$ triples exactly, reduce each fraction, and re-encode the observation.

We employ action masking based on the pruning methods described in Section \ref{subsec:newDivSel}. At each state we compute a Boolean mask over $\mathcal{P}_n\times\{\text{left},\text{right}\}$ where the only actions possible are the ones allowed by the pruning for the particular state. This reduces branching substantially while preserving optimal solutions under the assumptions of the heuristics.

Masked actions are treated as illegal by adding $-\infty$ to the corresponding logits before softmax during both training and inference. For the Clifford+T circuits, we also describe an efficient way to compute this mask in Section \ref{subsec:newDivSel}.

Episodes terminate upon reaching any permutation of the identity (success) or when a step cap is reached, which results in an unsuccessful final state.

\subsubsection{Policy/value network}
\label{sec:net}

Let $C=2\cdot 3B$ denote the input channel dimension. The policy/value network is intentionally compact:
\begin{enumerate}
  \item \textbf{Pointwise 2D convolution:} a $1{\times}1$ convolution mapping $\mathbb{R}^{4^n\times 4^n\times C}\to \mathbb{R}^{4^n\times 4^n\times L_1}$, acting as a learned per-entry embedding of the binary encodings from both views.
  \item \textbf{Flatten:} reshape to a vector of length $4^{2n}L_1$.
  \item \textbf{Shared linear block:} one fully-connected layer with nonlinearity.
  \item \textbf{Heads:} (i) an \emph{action head} producing logits over $2\cdot 4^n$ actions; (ii) a \emph{value head} producing a scalar $v\in\mathbb{R}$ estimating the episodic return $G$.
\end{enumerate}
Masked actions are implemented in practice by adding a large negative value to the corresponding logits before softmax.

\subsection{Model training}
Here we describe how the models are trained. We use an AlphaZero-style method adapted for single-player games, and combine this with curriculum learning where the targets are selected according to a given difficulty that grows as the training progresses.

\subsubsection{Single-player AlphaZero training}
\label{sec:az}

AlphaZero \cite{Silver2018AlphaZero} couples a neural network with Monte Carlo Tree Search (MCTS). The network, as described in Section \ref{sec:net}, takes a representation of a given state $s$ and outputs (i) a probability distribution over actions $l_\theta(s)$ that acts like a learned heuristic for where to search, and (ii) an estimate$ v_\theta(s)$ (value) of how good the state is; in our case, proportional to the distance to a Clifford. At a given state $s$, the MCTS builds a search tree rooted at $s$ where each node at level $l$ corresponds to a state reachable from $s$ by $l$ actions. The tree is constructed by running the following process repeatedly:

\begin{enumerate}
    \item Starting from the root, traverse the current tree until a leaf is reached by selecting an action at each step by selecting the max PUCT (policy-guided UCT) as  described in \cite{Silver2018AlphaZero}:
    \begin{equation}
U(s,a)=V(s,a)+c_{\mathrm{puct}}\,P_\theta(a|s)\frac{\sqrt{\sum_b N(s,b)}}{1+N(s,a)},
\end{equation}
where N(s,a) counts the number of visits so far to the node that results from applying action $a$ to $s$. 
    \item Estimate value and action probabilities for current leaf node with the neural network, and expand current branch by adding a leaf node by sampling an action from the estimated action distribution. Here, masked actions are ignored so they can never form part of the tree.
    \item Backtrack the estimated value and visit through the tree path nodes to update estimated values along the path. Unlike two-player win/loss backups, we propagate the \emph{undiscounted episodic return} $G\in[-1,1]$ derived from Eq.~\eqref{eq:reward}.
\end{enumerate} 

After a fixed budget of simulations, the visit counts and values at the first level of the tree are used to form improved probability estimates $\pi$ and returns $z$ that are in turn used to improve the network's predictions.

During the training, we generate multiple trajectories by preparing random initial states, and collect training tuples $(s, \pi, z)$, where $\pi$ is the MCTS-improved policy (visit counts normalized over unmasked actions) and $z=G\in[-1,1]$ is the terminal return of that episode. The network is optimized with
\begin{equation}
  \mathcal{L} \;=\; 
  \underbrace{\mathrm{CE}\!\left(\pi,\ \mathrm{softmax}\big(\ell_\theta(s)+\log m\big)\right)}_{\text{policy (masked)}}\ +\ 
  \lambda_v\,\underbrace{\big(v_\theta(s)-z\big)^2}_{\text{value}}\ \ 
\end{equation}
where $\ell_\theta$ are action logits, $m$ is the $0/1$ mask (adding $\log m$ enforces hard masking), $\mathrm{CE}$ denotes the cross entropy loss function and $\lambda_v$ is a hyperparameter.

\subsubsection{Curriculum learning.}

We train the model with a reverse construction curriculum. For difficulty $D$, we sample an initial channel $X_0=I$ and apply $D$ random actions from $\mathcal{P}_n\times\{\text{left},\text{right}\}$ to create the start state $X^{(D)}$. By construction, $X^{(D)}$ admits a solution with $\le D$ steps. During training at difficulty $D$, we cap the episode length at $D$ and mark failures accordingly. We start at $D=1$ and increment $D$ by one whenever the success rate at $D$ (under the training policy with MCTS; see below) exceeds a fixed threshold $\tau$.

The scalar reward is fixed and independent of difficulty:
\begin{equation}
  r_t \;=\;
  \begin{cases}
    \;\;\;\,1.0, & \text{if success state (terminal)},\\[2pt]
    -0.5, & \text{if unsuccessful final state (terminal)},\\[2pt]
    -\dfrac{0.5}{\texttt{max\_steps}}, & \text{otherwise.}
  \end{cases}
  \label{eq:reward}
\end{equation}
Here, \texttt{max\_steps} is a global cap on the number of steps that is constant throughout the training.

Hence the undiscounted episodic return $G=\sum_t r_t$ satisfies $G\in[-1,1]$. In particular, failure yields $-1<=G<=-0.5$, while a success in $T\!\le\!\texttt{max\_steps}$ steps yields $G=1-\frac{0.5(T-1)}{\texttt{max\_steps}}\in(0.5,\,1]$. This absolute scaling encourages the value head to estimate the \emph{distance-to-goal} uniformly across curriculum levels.

Training and evaluation are interleaved. When the success rate at difficulty $D$ exceeds $\tau$, we advance $D\!\leftarrow\! D\!+\!1$. This effectively mitigates the well-known sparse rewards problem when training RL models; since the model is fairly competent at difficulty $D$, in order to succeed at difficulty $D+1$ it is enough for the model to incrementally learn how to bring a $D+1$ state into a $D$ state that the model is already familiar with.

\subsection{Model inference and benchmarking}
\label{subsec:inferBenchmark}

Once a model has been trained up to a given difficulty $D$, we can use it to perform circuit synthesis by setting the initial state to the channel representation of the target unitary and unrolling the state trajectory step by step, retaining the sequence of selected gates as the circuit that implements the unitary (up to a Clifford). 

Each step of this unrolling proceeds as following:
\begin{enumerate}
    \item Take the current state (channel representation) and process it into an observation (as described in earlier sections).
    \item Compute the action masks with the pruning method described earlier.
    \item Estimate the probability of each of the allowed actions by applying the RL model to the observation with the action masks.
    \item Select an action from the allowed actions based on the distribution provided by the model.
    \item Evolve the state from the selected action, and stop if a Clifford is reached.
\end{enumerate}

Steps 3 and 4 can be done in different ways. For step 4, one can make the method deterministic by always picking the highest probability (i.e. greedy decoding) or generate alternative trajectories (and therefore different implementations) for the same target unitary by sampling from the distribution, potentially using a temperature parameter. When we do a greedy decoding we refer to our algorithm as \textbf{Greedy} and when we use $k$ samples, we refer to it as $\textbf{Sample}_{-}{\textbf{k}}$. For step 3, the straightforward way to do the probability estimation is to directly use the model outputs from the action head; however, since the model has been trained with AlphaZero and also provides a value head (an estimation of the distance to a Clifford at a given state), one could also generate an MCTS estimate of these probabilities as done in the training (at a much higher computational cost). 

From benchmarking we have observed that stochastic sampling without MCTS suffices and provides much faster synthesis times. We have also observed that a better way to trade computational effort for solution quality is to run $K$ independent rollouts by sampling and return the best (shortest) trajectory.

\subsection{A new algorithm for faster Divide and Select}
\label{subsec:newDivSel}

We have already mentioned that we use pruning techniques \cite{2021_MM, 2024_Mcs} in order to reduce the search space. Very briefly, it is a divide-and-select rule. At each level of the search tree the nodes are divided into two groups, depending upon the change in sde with respect to their parents. Unitaries in the set with the minimum cardinality become parents for the next level nodes, while the rest are discarded. We have observed before that in order to perform the pruning procedure, with each selected unitary (non-leaf node) we need to multiply all unitaries of the generating set. This can be computationally intensive. In this subsection we discuss some ways to reduce the number of multiplications at each node, for the specific case of Clifford+T set. 

Our intuition is as follows. In order to perform the Divide and Select operation at a particular node, we need to estimate the set of children unitaries for which the sde increases or non-increases. Instead of performing all the $4^n-1$ multiplications we attempt to identify the generating set unitaries i.e. $\chan{R(P)}$, that increases or non-increases the sde of the children (product) unitaries, using certain rules. After dividing the generating set unitaries, we select the smallest set, randomly select one generating set unitary from this set, and then multiply it with the parent unitary in order to get the child unitary. 

Suppose we have a $4^n\times 4^n$ matrix, $M_{Pauli}$, where each row and column corresponds to an $n$-qubit Pauli. This matrix stores information about product of Paulis and commutativity i.e. $M_{Pauli}[P_1,P_2] = [P_3, a]$ if $P_1P_2 = \pm i^a P_3$, where $a\in\{0,1\}$ and $P_1,P_2,P_3\in\pauli_n$. Here we assume that we encode the Paulis as integers. Let $\chan{U}$ be the parent unitary under consideration. $S_{col}$ is a $4^n$-length array that, for each column of $\chan{U}$, stores information about the position of the entries with $\sde_{\sqrt{2}}(\chan{U})$ (Equation \ref{eqn:sdeSqrt2M}). That is, if $S_{col}[j] = [i, k, \ell]$, then it implies that entries at positions $(i,j)$, $(k,j)$ and $(\ell,j)$ have the largest sde.  

Suppose we want to identify the generating set unitaries, $\chan{R(P)}$, that can increase or non-increase the sde of the child (or product) unitary, $\chan{U_p}$, after multiplication with $\chan{U}$. We know that half of the rows of $\chan{U}$ gets copied into $\chan{U_p}$. The other half of $\chan{U_p}$ is obtained by adding pairs of row of $\chan{U}$ and multiplying by $\frac{1}{\sqrt{2}}$ (refer \cite{2021_MM} and Appendix \ref{app:mult}). Let $u, v\in\intg\left[\frac{1}{\sqrt{2}}\right]$ such that they have the same sde. That is, they can be expressed in the form $u = \frac{a+b\sqrt{2}}{\sqrt{2}^k}$ and $v = \frac{c+d\sqrt{2}}{\sqrt{2}^k}$, where $a,c\in 2\intg+1$. From Fact 3 in \cite{2021_MM}, $\sde_{\sqrt{2}}\left(\frac{u\pm v}{\sqrt{2}}\right) = k$. Specifically, we have the following.
\begin{eqnarray}
   w &=& \frac{1}{\sqrt{2}}(u\pm v) = \frac{(a\pm c)+(b\pm d)\sqrt{2}}{\sqrt{2}^{k+1}}  \nonumber \\
   &=&\frac{1}{\sqrt{2}^k} \left( (b\pm d) + \frac{a\pm c}{2}\sqrt{2} \right) \qquad [a\pm c \in 2\intg]  
   \label{eqn:sdeSame}
\end{eqnarray}
If $(b\pm d) \in 2\intg+1 $ then we do no further reduction and $\sde_{\sqrt{2}}(w) = k$ i.e. sde remains unchanged. Else we do further reduction.
\begin{eqnarray}
    w &=& \frac{1}{\sqrt{2}^{k-1}} \left( \frac{a\pm c}{2} + \frac{b\pm d}{2}\sqrt{2}  \right)     \label{eqn:sdeDec}
\end{eqnarray}
In $\chan{U}$ if any column has a single max sde entry, say at (i,j), then sde increases for those $\chan{R(P)}$ that has $\frac{1}{\sqrt{2}}$ at the $i^{th}$ diagonal. This implies that for half of the $R(P)$s, sde will surely increase. So we need not consider those generating set unitaries. That is, we can avoid at least $4^n/2$ multiplications here. 

If a column has more than one 1 max sde entry then we do the following. Suppose $S_{col}[j][0]=i$. Then we consider those $\chan{R(P)}$ that has $\frac{1}{\sqrt{2}}$ at the $i^{th}$ diagonal. Let $M_{Pauli}[i,j]=[k,1]$ (remember they have to anti-commute). Then we consider $R(k)$. Now from the entries in $\chan{R(k)}$ we check if any max sde entry interacts with another non-max sde entry. This we can easily check from $S_{col}$. If it does then sde increases and we stop at this moment. Else sde non-increases. 

Hence, we save a lot by reducing the number of complete multiplications. More details about the above procedure, including pseudocodes have been provided in Appendix \ref{app:pseudocode}. Here we remark that if we want a finer division, say according to sde increase, decrease or unchanged, then we can verify further constraints on $a,b,c,d$, as in Equations \ref{eqn:sdeSame}, \ref{eqn:sdeDec}. Similar procedures can also be developed for the Clifford+CS gate set.
 \section{Discussion and Conclusion}
\label{sec:discuss}

In this paper we have investigated the potential of applying RL for the task of optimizing the T-count and CS-count while synthesizing quantum circuits for unitaries that are exactly implementable by the Clifford+T and Clifford+CS gate sets, respectively. This is the first ML algorithm that works with the channel representation of unitaries, that has a number of advantages, as explained in earlier sections. We have also appropriately designed our procedures in order to reduce the search space by adapting existing pruning heuristics and canonicalization of operators. Incorporation of these fundamental changes make our RL algorithm significantly faster, with a higher success rate and improvement factor (i.e. less non-Clifford count), as compared to previous RL algorithms. We are also able to synthesize circuits for unitaries with close to an order of magnitude higher T-count and gate-count compared to other RL methods such as the one described in \cite{2024_RDUetal}, and to the best of our knowledge the largest circuits of any other method so far (ML-based and otherwise). This is also the first ML algorithm that, to the best of our knowledge, explicitly optimizes non-Clifford gates for multi-qubit unitaries for generic universal gate sets. This work also highlights the importance of developing pruning heuristics for exhaustive search procedures, which have played a crucial role in achieving the reported performance. 

An interesting observation is that we could very easily match the performance of the known linear-time algorithms for 1-qubit Clifford+T and 2-qubit Clifford+CS, but the scaling was clearly worse for 2+ qubits and 3+ qubits for Clifford+T and Clifford+CS respectively. While this does not discard the existence of linear-time algorithms for higher qubit counts, it provides some additional numerical evidence that they may not exist. It is also an example on how RL can be used to probe the existence of efficient algorithms for where efficient proven solutions are not known.

Though the channel representation offers a host of advantages, one drawback is the fact that it maps a $2^n\times 2^n$ unitary to one with size $4^n\times 4^n$. For larger number of qubits this can impose a significant constraint on both the time and space complexity. In the future we intend to probe further in order to compensate this disadvantage for the synthesis of even larger unitaries, while optimizing the non-Clifford count. 

In this paper we have demonstrated how optimal unitary synthesis algorithms can help in significantly reducing the asymptotic gate complexity of larger unitaries. For the specific cases of controlled cyclic shift, controlled adder and multiplier circuits, to the best of our knowledge, we are not aware of any algorithm (synthesis or resynthesis) that is able to show such asymptotic reduction. This illustrates that our algorithms and the results derived in this paper can be useful for resource estimates of quantum algorithms. Unitary synthesis algorithms can also be used along with resynthesis algorithms as a first or second pass of optimization. The extent of optimizations possible with such mixed procedures and the trade-offs have been left for future study.

\section*{Author contributions}

D.K. designed and implemented the RL algorithms. The non-RL concepts were developed and implemented by A.J-A and P.M. All the authors contributed equally in the preparation of the manuscript.

\section*{Competing interests}

The authors have no competing interests to declare.

\section*{Code and Data availability}

The RL environments are available as part of Qiskit-Gym (\href{https://github.com/AI4quantum/qiskit-gym}{https://github.com/AI4quantum/qiskit-gym}).

\bibliographystyle{unsrt}
\bibliography{RL_ref}

\appendix

\section{Some additional preliminaries}
\label{app:prelim}

\subsection{Cliffords and Paulis}
\label{app:clifford}

The \emph{single qubit Pauli matrices} are as follows:
\begin{eqnarray}
 \X=\begin{bmatrix}
     0 & 1 \\
    1 & 0
    \end{bmatrix} \qquad  
 \Y=\begin{bmatrix}
     0 & -i \\
     i & 0
    \end{bmatrix} \qquad 
 \Z=\begin{bmatrix}
     1 & 0 \\
     0 & -1
    \end{bmatrix}\nonumber
\label{eqn:Pauli1}
\end{eqnarray}

The \emph{$n$-qubit Pauli operators} are :
$
 \pauli_n=\{Q_1\otimes Q_2\otimes\ldots\otimes Q_n:Q_i\in\{\id,\X,\Y,\Z\} \}.
$

The \emph{single-qubit Clifford group} $\cliff_1$ is generated by the Hadamard and phase gates :
$
 \cliff_1=\braket{\had,\phase} 
 $
where
\begin{eqnarray}
 \had=\frac{1}{\sqrt{2}}\begin{bmatrix}
       1 & 1 \\
       1 & -1
      \end{bmatrix}\qquad 
 \phase=\begin{bmatrix}
       1 & 0 \\
       0 & i
      \end{bmatrix}\nonumber
\end{eqnarray}
When $n>1$ the \emph{$n$-qubit Clifford group} $\cliff_n$ is generated by these two gates (acting on any of the $n$ qubits) along with the two-qubit $\CNOT=\ket{0}\bra{0}\otimes\id+\ket{1}\bra{1}\otimes\X$ gate (acting on any pair of qubits).

\section{Circuit construction for generating set unitaries}
\label{app:cktGen}

In this section we describe various methods to synthesize the circuits for each generating set unitary.

\paragraph{Circuit construction for $R(P)$ : } We can write $R(P) = \frac{1}{2}\left(1+e^{\frac{i\pi}{4}}\right)\id + \frac{1}{2}\left(1-e^{\frac{i\pi}{4}}\right)C\Z_{(q_i)}C^{\dagger} = C\T_{(q_i)}C^{\dagger} $, where $C\in\cliff_n$ such that $C\Z_{(q_i)}C^{\dagger} = P$.
In order to construct a circuit for unitary $R(P)$ we first find a Clifford $C\in\cliff_n$ such that $CPC^{\dagger} = \Z_{(q)}$, where $q$ is the $q^{th}$ qubit. This is equivalent to finding the circuit of a Clifford operator that diagonalizes the Pauli $P$. We can use the (non-ML) algorithm in \cite{2020_dBT} or the RL algorithm in \cite{2025_DKMetal}. Thus a circuit for $R(P)$ consists of $\T_{(q)}$, conjugated by the Clifford $C$.

\paragraph{Circuit construction for $G_{P_1,P_2}$ \cite{2024_Mcs} : } We can construct a circuit implementing $G_{P_1,P_2}$ by deriving the conjugating Clifford and determining the control and target qubit of the CS gate. Given a pair of commuting non-identity Paulis $P_1,P_2$, we use the algorithm in \cite{2020_dBT} or \cite{2025_DKMetal} in order to derive a conjugating Clifford $C'\in\cliff_n$ such that $C'P_1C'^{\dagger}=P_1'$ and $C'P_2C'^{\dagger}=P_2'$, where $P_1'=\bigotimes_{j=1}^nQ_j$, $P_2'=\bigotimes_{j=1}^nR_j$ and $Q_j,R_j\in\{\id,\Z\}$. That is, the output of this algorithm is a pair of $\Z$-operators i.e. $n$-qubit Paulis that are tensor product of either $\id$ or $\Z$. Then we use the following conjugation relations,
\begin{eqnarray}
  &&  \swap (\id\otimes\Z)\swap=\Z\otimes\id;\quad \CNOT_{\q{j;k}} (\id_{\q{j}}\otimes\Z_{\q{k}})\CNOT_{\q{j;k}}=\Z_{\q{j}}\otimes\Z_{\q{k}}    \nonumber \\
   && \CNOT_{\q{j;k}} (\Z_{\q{j}}\otimes\id_{\q{k}})\CNOT_{\q{j;k}}=\Z_{\q{j}}\otimes\id_{\q{k}}; \nonumber
\end{eqnarray}
in order to derive Clifford $C''\in\cliff_n$ such that $C''P_1'C''^{\dagger}=\Z_{\q{a}}$ and $C''P_2'C''^{\dagger}=\Z_{\q{b}}$, where $1\leq a,b\leq n$ and $a\neq b$. If $C=C'C''$, then we get $C'C''\Z_{\q{a}}C''^{\dagger}C'^{\dagger}=P_1$ and $C'C''\Z_{\q{b}}C''^{\dagger}C'^{\dagger}=P_2$. Thus a circuit for $G_{P_1,P_2}$ consists of $\cs_{\q{a,b}}$, conjugated by Clifford $C=C'C''$.

The complexity of synthesizing each of the generating set unitaries primarily depend on the complexity of the diagonalizing algorithms. Both the algorithms \cite{2020_dBT, 2025_DKMetal} have complexity that is polynomial in the number of qubits, specifically $O(n^2)$, where $n$ is the number of qubits. Hence the complexity of synthesizing each of the generating set unitaries is $O(n^2)$. Thus, in our context we can assume that each of these generating set unitaries can be synthesized efficiently.

\section{Multiplication by generating set unitaries}
\label{app:mult}

In this section we briefly describe the algorithms to multiply any matrix with the generating set unitaries. More details and pseudocodes can be found in the cited references.

\paragraph{Multiplication of $\chan{R(P)}$  \cite{2021_MM} } :
Let $U_p=R(P)U$ where $U$ is another $2^{2n}\times 2^{2n}$ matrix. Then,
\begin{eqnarray} 
 U_p[r,j] = \sum_{k=1}^{2^{2n}} \chan{R(P)}[r,k]U[k,j].    \nonumber
 \label{eqn:mult}
\end{eqnarray}

\begin{enumerate}
    \item Let $\chan{R(P)}[r,r] = 1$, implying $\chan{R(P)}[r,s]\neq 0$, for each $r\neq s$. Then, $U_p[r,j] = U[r,j]$, for each $j\in \{1,\ldots,2^{2n}\}$ and so $U_p[r,.]\leftarrow U[r,.]$ i.e. the $r^{th}$ row of $U$ gets copied into the $r^{th}$ row of $U_p$.

    \item Let $\chan{R(P)}[r,r]=\frac{1}{\sqrt{2}}$. From \cite{2021_MM} there exists an off-diagonal element $s$ such that $\chan{R(P)}[r,s] = \pm\frac{1}{\sqrt{2}}$ and the remaining entries in that row are 0. This implies that 
    \begin{eqnarray}
        U_p[r,j] = \frac{1}{\sqrt{2}} \left(U[r,j]\pm U[s,j]\right), \nonumber
    \end{eqnarray}
    equivalently $U_p[r,.]\leftarrow\frac{1}{\sqrt{2}}\left(U[r,.]\pm U[s,.]\right)$.
\end{enumerate}

\paragraph{Multiplication by $\chan{G_{P_1,P_2}}$ \cite{2024_Mcs} } :
Let $U_p'=\chan{G_{P_1,P_2}}U'$, where $U'$ is another matrix of dimension $2^{2n}\times 2^{2n}$. Then,
\begin{eqnarray}
    U_p'[r,j] = \sum_{k=1}^{2^{2n}} \chan{G_{P_1,P_2}}[r,k] U'[k,j]   \nonumber
\end{eqnarray}
and it can be computed very efficiently with the following observations.

\begin{enumerate}
    \item Suppose $\chan{G_{P_1,P_2}}[r,r]=1$, implying $\chan{G_{P_1,P_2}}[r,s]=0$ for each $s\neq r$. Then,
    \begin{eqnarray}
        U_p'[r,j]=\chan{G_{P_1,P_2}}[r,r]U'[r,j]=U'[r,j]  \qquad [\forall j\in\{1,\ldots,2^{2n}\} ] \nonumber
    \end{eqnarray}
and so $U_p'[r,.]\leftarrow U'[r,.]$ i.e. the $r^{th}$ row of $U'$ gets copied into the $r^{th}$ row of $U_p'$. 

    \item Let $\chan{G_{P_1,P_2}}[r,r]=\frac{1}{2}$. From \cite{2024_Mcs}, we know there are 3 other non-zero off-diagonal elements. Let $\chan{G_{P_1,P_2}}[r,s]=\pm\frac{1}{2}$, $\chan{G_{P_1,P_2}}[r,s']=\pm\frac{1}{2}$ and $\chan{G_{P_1,P_2}}[r,s'']=\pm\frac{1}{2}$.
\begin{eqnarray}
    U_p'[r,j]&=&\chan{G_{P_1,P_2}}[r,r]U'[r,j]+\chan{G_{P_1,P_2}}[r,s]U'[s,j]+\chan{G_{P_1,P_2}}[r,s']U'[s',j]+\chan{G_{P_1,P_2}}[r,s'']U'[s'',j] \nonumber \\
     &=&\frac{1}{2}\left( U'[r,j]\pm U'[s,j]\pm U'[s',j]\pm U'[s'',j] \right)   \nonumber
\end{eqnarray}
Thus we see that the $r^{th}$ row of $U_p'$ is a linear combination of the $r^{th}, s^{th}, s'^{th}, s''^{th}$ rows of $U'$, multiplied by $\frac{1}{2}$, i.e.
$ U_p'[r,.]\leftarrow\frac{1}{2}\left( U'[r,.]\pm U'[s,.]\pm U'[s',.]\pm U'[s'',.] \right) $.
\end{enumerate}

\section{Pseudocode}
\label{app:pseudocode}

In this section we provide more details and pseudocode for the implementation of the Divide and Select procedure described in Section \ref{subsec:newDivSel}. We use the compact representation of $\chan{R(P)}$ as an array of indices, as described in Section 3.4 of \cite{2021_MM}. We assume each $n$-qubit Pauli is encoded by an integer. Additionally, we define the following data structures.

(a) $M_{Pauli}$ is a $4^n\times 4^n$ matrix where each row and column is an $n$-qubit Pauli. If $P_1P_2 = \pm i^{a} P_3$, where $a \in \{0,1\}$, then $M_{Pauli}[P_1,P_2] = [P_3, a]$ i.e. it stores the product Pauli and the commutation information. If Paulis commute then $a = 0$, else it is $1$. 

(b) $S_{col}$ is an array of size $4^n$, corresponding to the columns in each channel representation matrix. In this array we store information about the position of max sde entries in each column. So each entry of this array is a set of row indices. Suppose $S_{col}[j] = [i, k, \ell]$, this implies we have max sde entries at positions $(i,j)$, $(k,j)$ and $(\ell,j)$.

(c) We keep another array $S_{col,1}$ which stores indices of those columns which have only one max sde.

(d) $R_{mult}$ is an array of size $4^n-1$, corresponding to the number of generating set unitaries i.e. $R(P)$. This is initialized to all 0 before each node is considered. Every time we consider multiplication by $R(P)$, we set $R_{mult}[P]$ to 1. 

If we want to keep track how the sde changed then we can allot more values, say 0 implies not considered, 1 implies sde decrease, 2 implies sde same and 3 implies sde increase. It will be good to have copies of this array for each node. Then after the selection, we know how to obtain the next level nodes. 

If we want to divide into 3 groups according to change in sde - inc, same, dec, then we need to also keep track of the change in the second highest sde if and only if this second highest sde is only one integer less than the max sde. For this, we define the following additional data structures.

(e) $S_{col,1}'$ : This array stores the indices of those columns which have only one entry with (max -1) sde.

(f) $S_{col}'$ : This array has size $4^n$, corresponding to the number of columns in each channel representation matrix. $S_{col}'[j]$ stores information about those entries in the $j^{th}$ column that has sde (max - 1). So each entry of this array is a set of row indices. Suppose $S_{col}'[j] = [i, k, \ell]$, this implies we have (max-1) sde entries at positions $(i,j)$, $(k,j)$ and $(\ell,j)$.

\begin{algorithm}
\scriptsize
\caption{DIV-N-SEL}

\KwIn{(i) $\chan{U}$; (ii) $\mathcal{A} = \{\mathcal{A}_P : P\in\pauli_n  \}$ }

\KwOut{Number of increase, decrease or same}

$inc = 0$; $dec = 0$; $same = 0$ \;

\For{each $i$ in $len(S_{col,1})$ }
{
    $j = S_{col,1}[i]$  \;
    \For{each $k$ in  $1,\ldots 4^n-1$ }
    {
        \If{$M_{Pauli}[j,k][1] == 1$ \tcp*[h]{Check if anti-commute} }
        {
            $\ell = M_{Pauli}[j,k][0]$  \;
            \If{$R_{mult}[\ell] == 0$  \tcp*[h]{If not already considered}}
            {
                $R_{mult}[\ell] = 3$    \;
                $inc = inc+1$   \;
            }
        }
    }
}

\For{each $j = 1,\ldots,4^n-1 $}
{
    \If{$len(S_{col}[j]) > 1$}
    {
        \For{each $i\in len(S_{col}[j])$}
        {
            \If{$M_{Pauli}[i,j][1]==1$} 
            {
                $\ell = M_{Pauli}[i,j][0]$  \;
                \If{$R_{mult}[\ell] == 0$ \tcp*[h]{If not already considered}}
                {
                    \For{each $(x,\pm y)\in \mathcal{A}_P$}
                    {
                        \If{($x \in S_{col}[j], y\notin S_{col}[j]$) or ($y \in S_{col}[j], x\notin S_{col}[j]$}
                        {
                            $R_{mult}[\ell] = 3$; $inc = inc+1$   \;
                            \textbf{break}  \;
                        }
                        If $x, y\in S_{col}[j]$ then change $R_{mult}[\ell]$ to 1 or 2 if initially it was 0. \tcp*[h]{Do this if divide into 2 groups - inc and non-inc}   \;
                        
                        \If{ $x, y\in S_{col}[j]$ \tcp*[h]{(Do this if divide into 3 groups - inc, same, dec)} }
                        {
                            Check change in sde using entries in $\chan{U}[x,j]$ and $\chan{U}[y,j]$ using Equations \ref{eqn:sdeDec}, \ref{eqn:sdeSame}   \;
                            If sde same then change $R_{mult}[\ell]$ to 2 if it was initially not 3  \;
                            If sde decrease then change $R_{mult}[\ell]$ to 1 only if it was initially 0 or 1   \;
                        }
                    }
                    Increase dec or same by 1 if $R_{mult}[\ell]$ is 1 or 2, respectively \;
                }
            }
        }
    }
}
If any $R_{mult}[\ell] = 0$ i.e. not considered then change it to 2 i.e. sde of matrix remains same    \;
If dividing into 3 groups then call CHANGE-SECOND-MAX   \;

\end{algorithm}

\begin{algorithm}
\scriptsize
\caption{CHANGE-SECOND-MAX}

\KwIn{(i) $\chan{U}$; (ii) $\mathcal{A} = \{\mathcal{A}_P : P\in\pauli_n  \}$ }

\KwOut{Number of increase, decrease or same}

\For{each $i$ in $len(S_{col,1}')$ }
{
    $j = S_{col,1}'[i]$  \;
    \For{each $k$ in  $1,\ldots 4^n-1$ }
    {
        \If{$M_{Pauli}[j,k][1] == 1$ \tcp*[h]{Check if anti-commute} }
        {
            $\ell = M_{Pauli}[j,k][0]$  \;
            \If{ $R_{mult}[\ell] != 3$ }
            {
                $R_{mult}[\ell] = 2$    \;    
                same = same + 1   \;
            }
        }
    }
}

\For{each $j = 1,\ldots,4^n-1 $}
{
    \If{$len(S_{col}'[j]) > 1$}
    {
        \For{each $i\in len(S_{col}'[j])$}
        {
            \If{$M_{Pauli}[i,j][1]==1$} 
            {
                $\ell = M_{Pauli}[i,j][0]$  \;
                \If{$R_{mult}[\ell] != 3$}
                {
                    \For{each $(x,\pm y)\in \mathcal{A}_P$}
                    {
                        \If{($x \in S_{col}'[j], y\notin S_{col}'[j]$) or ($y \in S_{col}'[j], x\notin S_{col}'[j]$}
                        {
                            $R_{mult}[\ell] = 2$; $same = same+1$   \;
                            \textbf{break}  \;
                        }
                    }
                }
            }
        }
    }
}

\end{algorithm}

\end{document}